\documentclass[preprint,12pt]{elsarticle}
\usepackage{epsfig}
\usepackage{graphicx}
\usepackage[ansinew]{inputenc}
\usepackage{epstopdf}
\usepackage{psfrag}
\usepackage{amsmath}
\usepackage{amssymb}
\usepackage{float}
\usepackage{multirow}
\usepackage{lineno}
\usepackage{setspace}
\usepackage{flushend}
\usepackage{array}
\usepackage{styles/tabu}
\usepackage{bm}
\usepackage{color}
\usepackage{url}
\usepackage{natbib}
\usepackage{fixltx2e}
\newcolumntype{L}[1]{>{\raggedright\let\newline\\\arraybackslash\hspace{0pt}}m{#1}}
\newcolumntype{C}[1]{>{\centering\let\newline\\\arraybackslash\hspace{0pt}}m{#1}}
\newcolumntype{R}[1]{>{\raggedleft\let\newline\\\arraybackslash\hspace{0pt}}m{#1}}
\usepackage[figuresright]{rotating}
\usepackage{subfigure}
\usepackage[title,titletoc,toc]{appendix}
\usepackage{mathrsfs}
\usepackage{amsfonts}
\usepackage{svgcolor}
\usepackage{styles/ieeetrantools}
\usepackage{makecell}
\usepackage{booktabs}

\usepackage{xcolor}
\definecolor{verde}{rgb}{0.0, 0.5, 0.0}
\definecolor{rojo}{rgb}{0.7, 0, 0.0}
\definecolor{grisClaro}{rgb}{0.65, 0.65, 0.65}

\def \tinyREF   {{\scriptscriptstyle \! r\!e\!f}}
\def \tinyREFVAP   {{\scriptscriptstyle \! r\!e\!f_{\!v}}}

\def \tinyREFTWO   {{\scriptscriptstyle \! r\!e\!f_{\!2}}}
\def \tinySEC   {{\scriptscriptstyle \! s\!e\!c}}
\def \tinyPCM   {{\scriptscriptstyle \! p\!c\!m}}

\def \tinyPCMSOL   {{\scriptscriptstyle \! p\!c\!m_{\!s}}}
\def \tinyPCMLIQ   {{\scriptscriptstyle \! p\!c\!m_{\!l}}}

\def \tinyMAX   {{\scriptscriptstyle \! m\!a\!x}}
\def \tinyMIN   {{\scriptscriptstyle \! m\!i\!n}}

\def \tinyINT    {{\scriptscriptstyle \!i\!n\!t}}
\def \tinyEXT    {{\scriptscriptstyle \!e\!x\!t}}
\def \tinyIN    {{\scriptscriptstyle i\!n}}
\def \tinyOUT    {{\scriptscriptstyle o\!u\!t}}

\def \tinyCOND    {{\scriptscriptstyle c\!o\!n\!d}}
\def \tinyCONV    {{\scriptscriptstyle c\!o\!n\!v}}
\def \tinyCONVINT    {{\scriptscriptstyle c\!o\!n\!v\!,i\!n\!t}}

\def \tinyCONDWALL    {{\scriptscriptstyle c\!o\!n\!d\!,w\!a\!l\!l}}

\def \tinyCONVEXT    {{\scriptscriptstyle c\!o\!n\!v\!,e\!x\!t}}

\def \tinyWALL   {{\scriptscriptstyle w\!a\!l\!l}}
\def \tinyLAT   {{\scriptscriptstyle l\!a\!t}}
\def \tinyLATMIN    {{\scriptscriptstyle \!l\!a\!t\!-}}
\def \tinyLATMAX    {{\scriptscriptstyle \!l\!a\!t\!+}}

\def \tinyLAY {{\scriptscriptstyle \! l\!a\!y}}
\def \tinyNLAY {{n_\tinyLAY}}
\def \tinyPCMOne   {{\scriptscriptstyle \! p\!c\!m\!,1}}

\def \tinyPCMK   {{\scriptscriptstyle \! p\!c\!m\!,k}}
\def \tinyPCMKMenos {{\scriptscriptstyle \! p\!c\!m\!,k-1}}
\def \tinyPCMKMas {{\scriptscriptstyle \! p\!c\!m\!,k+1}}
\def \tinyPCMNLAY   {{\scriptscriptstyle \! p\!c\!m\!,\tinyNLAY}}
\def \tinyPCMKCero {{\scriptscriptstyle \! p\!c\!m\!,k_{0}}}
\def \tinyPCMKCeroMenos {{\scriptscriptstyle \! p\!c\!m\!,k_{0}\!-1}}
\def \tinyPCMKCeroMenosJ {{\scriptscriptstyle \! p\!c\!m\!,k_{0}\!-j}}
\def \tinyKCero {{\scriptscriptstyle \! k_{0}}}
\def \tinyKCeroMenos {{\scriptscriptstyle \! k_{0}\!-1}}

\def \tinyKCeroMenosI {{\scriptscriptstyle \!k_{0}\!-i}}

\journal{Applied Thermal Engineering}

\begin{document}

\bstctlcite{bibliografia:BSTcontrol}

\begin{frontmatter}

\title{Efficient simulation strategy for PCM-based cold-energy storage systems\footnote{© 2018. This manuscript version is made available under the CC-BY-NC-ND 4.0 license \url{https://creativecommons.org/licenses/by-nc-nd/4.0/}. The link to the formal publication is \url{https://doi.org/10.1016/j.applthermaleng.2018.05.008}}}

\author[us]{Guillermo Bejarano}

\author[us]{Manuel Vargas\corref{mycorrespondingauthor}}
\cortext[mycorrespondingauthor]{Corresponding author}
\ead{mvargas@us.es}

\author[us]{Manuel G. Ortega}
\author[us]{Fernando Casta{\~n}o}    
\author[ufsc]{Julio E. Normey-Rico}

\address[us]{Department of Systems Engineering and Automation, University of Seville, Seville, Spain}
\address[ufsc]{Department of Automation and Systems (DAS), Federal University of Santa Catarina, Florianópolis, SC, Brazil}

\begin{abstract}
{
	\emph{This paper proposes a computationally efficient simulation strategy for cold thermal energy storage (TES) systems based on phase change material (PCM). Taking as a starting point the recent design of a TES system based on PCM, designed to complement a vapour-compression refrigeration plant, the new highly efficient modelling strategy is described and its performance is compared against the pre-existing one. The need for a new computationally efficient approach comes from the fact that, in the near future, such a TES model is intended to be used in combination with the model of the own mother refrigeration plant, in order to address efficient, long-term energy management strategies, where computation time will become a major issue. Comparative simulations show that the proposed computationally efficient strategy reduces the simulation time to a small fraction of the original figure (from around 1/30th till around 1/120th, depending on the particular choice of the main sampling interval), at the expense of affordable inaccuracy in terms of the PCM charge ratio.}
}
\end{abstract}

\begin{keyword}

    Refrigeration system 	\sep
    Cold-energy storage  	\sep
    Phase change materials  \sep
    Dynamic modelling 		\sep
    Computational efficiency

\end{keyword}

\end{frontmatter}


\section{Introduction} \label{Introduction}

Cold-energy production via vapour-compression systems is definitively the most common method used worldwide. Significant efforts to increment energy efficiency while reducing environmental impact of current vapour-compression systems have been carried out in recent years. A novel line of research focuses not just on efficient cold-energy generation, but also on cold-energy management, including thermal energy storage systems (TES). The main idea is to use a certain reservoir to manage cold energy, in such a way that it can be stored and released according to the needs at any given time. This strategy allows to streamline the design stage, helping to avoid any undesirable oversizing of the system, in order to satisfy peak demand periods. Thereby, the equipment can be more efficiently used and energy consumption can be reduced \cite{dincer2002bthermal,maccracken2004thermal}.

From an economic point of view, TES systems enable the scheduling of cold-energy production, so that the consumer benefits from low-price time slots, typically, when global demand is lower (\emph{peak-shifting}) \cite{rismanchi2012energy}. Several works in the literature address management of TES systems, following different control strategies \cite{wang2007cnovel,mosaffa2014advanced,shafiei2014model}.

Regarding the design of the cold-energy reservoir, storage tanks filled with phase change material (PCM) have become a successful trend over sensible-heat materials, due to their convenient thermodynamic properties for heat transfer. Not only their heat capacity is a very relevant factor, but also the fact that their temperature does not vary significantly, provided that the material remains in latent zone, which boosts heat transfer \cite{mehling2008heat}. A number of solid-liquid phase change materials designed for cold-energy storage applications can be found in the literature, including commercial and developing solutions \cite{mehling2008heat, oro2012review,sharma2009review}.

Different technologies regarding latent heat TES based on PCM have been proposed and experimentally tested in the literature, where a variety of geometries and fluid arrangements can be found \cite{oro2012review,verma2008review,dutil2011review}. Packed bed technology is among the most common configurations, where a certain volume includes a large number of small PCM nodules. First, cold heat transfer fluid (HTF) is used to charge the system by flowing through the volume and solidifying the PCM nodules. Afterwards, warm HTF circulates while melting the PCM nodules to discharge the system, which releases the previously stored cold energy \cite{saitoh1986high}.

Note that the same HTF is used within the packed bed technology to charge and discharge the system. However, in the application proposed in this work, the cold-energy storage tank is projected to complement an existing vapour-compression refrigeration facility. Then, the simultaneous operation of the refrigeration cycle is intended to charge the TES while satisfying the cooling load, in such a way that the cold HTF would correspond to the refrigerant during the charging stage. Furthermore, the objective is to remove heat from a secondary fluid, so that the warm HTF would not correspond to the refrigerant, but to another secondary fluid. This key difference with respect to the standard packed bed technology implies that an original hybrid structure has to be considered, where two different HTFs are involved. Besides, the tank is filled with PCM nodules, bathed in a certain liquid, called \emph{intermediate fluid}, while two bundles of pipes, corresponding to the refrigerant and the secondary fluid, run through the tank, being also dipped in the intermediate fluid (the constructive technical details can be found in a recent work by Bejarano \emph{et al.} \cite{Bejarano2017NovelSchemePCM}).

Concerning modelling, much effort has been devoted to packed bed technology \cite{zhang2001general,benmansour2006experimental,cheralathan2007effect,bedecarrats2009astudy}. The problem of predicting the behaviour of PCM-based systems is difficult to solve, given its non-linear nature. Moreover, there are moving interfaces whose displacement is determined by the latent heat stored or released at the boundary. Therefore, the position and velocity of those boundaries are not known \emph{a priori}.

There are two main approaches to model the behaviour of PCM-based systems. The first one relies on analytical models, using first-principle equations, whereas the second one involves numerical finite-element methods. Some relevant works of both modelling strategies are mentioned next.

On the one hand, regarding the analytical models, B{\'e}d{\'e}carrats \emph{et al.} study the behaviour of a test plant, consisting of a cylindrical tank containing spherical nodules, filled with phase change material, through which a liquid fluid flows for latent heat storage purpose \cite{bedecarrats1996phase,bedecarrats2009bstudy}. The tank is divided in several control volumes and the PCM nodules are considered as exchangers, in such a way that the transferred energy flux is proportional to the temperature difference between the fluid and the spherical nodule. The modelling of the heat flux transferred by each nodule considers a pure-conduction problem, as well as the dynamic evolution of the radius of the spherical boundary between the solid and liquid PCM phases inside the capsule. It is stated that, in the discharge, an \emph{effective} thermal conductivity of the liquid phase must be considered to model internal convection during melting. It is concluded that the simplified model considering the nodules as exchangers confirms the experimental results with reasonable accuracy.

Alternatively, Ismail and Moraes present a quantitative study of the solidification of the PCM enclosed in a spherical shell \cite{ismail2009numerical}. The mathematical model is also based on pure conduction in the PCM, subject to the boundary condition of constant surface temperature. The model is validated experimentally and the agreement is found to be satisfactory.

Furthermore, Amin \emph{et al.} experimentally study the freezing and melting of PCM encapsulated in a sphere \cite{amin2014effective}. They confirm empirically that the pure-conduction model is realistic when charging the PCM spheres, since the buoyancy forces have a negligible impact on the freezing process. However, as stated by other authors, it is necessary to consider an \emph{effective} thermal conductivity of the liquid phase during the discharge to retain the spherical geometry.

Recently, Bejarano \emph{et al.} have presented a continuous model of a novel TES system combined with a refrigeration cycle based on PCM, using the configuration previously described \cite{Bejarano2017NovelSchemePCM}. The modelling of the freezing/melting processes is based on the previous works but adapted to the novel setup, which is thoroughly described in that paper.

The analytical modelling approach provides a computationally efficient representation of the system dynamics, since the proposed models involve solving low-order differential equations. Nevertheless, as stated in the mentioned work by Bejarano \emph{et al.}, the continuous modelling involving a single inward freezing/melting front within the PCM capsule provides a limited description of the system dynamics: only strict full charging/discharging operations can be simulated \cite{Bejarano2017NovelSchemePCM}. If a realistic representation of the dynamic evolution of the system during any series of partial charging/discharging operations was intended, in such a way that an arbitrary number of moving freezing/melting boundaries could be present at the same time inside the PCM capsules, this would mean a potentially infinite-dimensional state vector. However, keeping in mind that the TES model is intended to be used within a more complex model including the refrigeration cycle, from which efficient energy management strategies are intended to be designed, partial charging/discharging operations dictated by economic and energy efficiency criteria seem likely to be scheduled. Thus, this intrinsic limitation of the analytical models represents a hard-to-overcome drawback.

On the other hand, regarding finite-element methods, Computational Fluid Dynamics (CFD) represent a very common alternative to model the melting and freezing processes taking place in each individual PCM nodule. CFD software packages such as {\sc gambit} or {\sc ansys fluent} are common examples in this context \cite{macphee2009thermodynamic}. 

To keep a reasonable representation of the physical processes taking place at the melting front, model grid density must be high enough to smoothly cover the solid–liquid interface. Diverse mathematical solutions have been studied in the literature \cite{dutil2011review, LIU2014659}. The fixed grid approach is able to deal with strong nonlinearities, not requiring explicit treatment of conditions on the phase change boundary, which allows to use standard solution procedures for the energy equations. However, the required high grid density is not needed elsewhere in the numerical domain. Therefore, adaptive grid or front tracking schemes evaluate the exact location of the moving boundary on a grid at each step \cite{ismail2002numerical}. There are two main approaches: the interface-fitting grids and the variable-space grids. In the interface-fitting grids (also referred to as variable time step methods), a uniform spatial grid but a non-uniform time step are used. Furthermore, in the variable-space grids, or dynamic grids, the number of time intervals is constant, while the spatial intervals are adjusted in such a way so that the moving boundary lies on a particular grid point, thus the spatial intervals are a function of time. Concerning computational load, it has been widely documented that CFD methods, in general, demand considerable computational resources, being more useful during design stages than to perform continuous simulations  over long periods of time, typically inherent to energy management strategies.

The authors have proposed a discrete model of the TES system combined with  refrigeration cycle in the mentioned recent work, similar to finite-element methods, which has been shown to provide a suitable description of the system dynamics for both, full charging/discharging cycles and any series of partial charging/discharging operations \cite{Bejarano2017NovelSchemePCM}. It is in fact a double-discrete model, since not only it is a time-discrete one, but also a volume quantisation of the PCM nodules is imposed, as the PCM capsules are conceptually divided in a given set of spherical layers. Albeit far below from the expected computing workload of classical CFD implementations, this discrete model is still computationally demanding, apart from the fact that the number of spherical layers represents a trade-off between reduction in modelling errors, with respect to the pure-conduction continuous model presented in the same work, and computational load.

Being aware that the TES model is intended to be used as a component of a global model comprising the refrigeration cycle, from which energy management strategies are to be performed, the computational load must be low enough. In fact, at the top level, where efficient energy management is the main objective, the simulated time scales might be in the order of weeks or months. Considering these time scales, nor the classical CFD methods nor the proposed discrete model provide reasonable simulation times. It justifies the need for an "accelerated" model which allows to simulate long time periods and design energy management strategies based on economic and efficiency criteria, always preserving reasonable accuracy.

In this paper, a highly time-efficient discrete model is presented, based on that one proposed by Bejarano \emph{et al.} \cite{Bejarano2017NovelSchemePCM}, which achieves a significant reduction in the simulation time, while preserving the overall accuracy in the description of the TES tank behaviour. The system dynamics are analysed and the fastest (related to the layers in sensible zone) are considered as negligible, which allows to cluster the external consecutive layers in sensible zone into a single thermal-conduction layer. According to this, cold energy can be assumed to be transferred at a constant rate during a certain time interval, that needs to be estimated. This analysis leads to an adaptive variable-step integration procedure, which allows to drastically reduce the simulation time by considering longer time steps and subdividing them into several time intervals to describe the changes caused by the transition from latent to sensible zone of every layer. The proposed variable-step integration procedure is not only valid for this specific discrete model, but it can be applied with minor modifications to other PCM-based systems.

The paper is organised as follows. Section \ref{secSystemDescription} concisely describes the TES system under study, which supports a pre-existing refrigeration facility. Section \ref{secDiscreteModel} shortly describes the discrete model used as a starting point, highlighting their specificities. The main contribution of this paper, regarding computationally efficient simulation, is addressed in Section \ref{secEfficientDiscreteModels}, where the new highly efficient discrete model is proposed, allowing a radical reduction in the required simulation time. Several comparative simulations of full charge/discharge and sequences of partial charging/discharging operations are provided in Section \ref{secAccuracyAssessment}, while computational efficiency of the proposed method is evaluated in Section \ref{secComputationalCost}. Eventually, the main conclusions are summarised in Section
\ref{secConclusions}.


\section{System description} \label{secSystemDescription}

\subsection{Notation}

Tables \ref{tabSymbols} and \ref{tabSubscriptsSuperscripts} provide the list of symbols and the subscript/superscript notation followed along the modelling description of the thermal storage system. Fluid properties, such as $c_{\!p}$, $\rho$, etc. are not constant, but dependent on temperatures, pressures, etc. Their instantaneous values are rigorously considered in the simulation models, provided by the \emph{CoolProp} tool \cite{CoolProp}.

\begin{table}
    \centering
    \caption{List of symbols}
    \label{tabSymbols}
    \scalebox{0.7}[0.7]{ \tabulinesep=0.5mm
        \begin{tabu} { L{1.3cm} L{10cm} L{2cm} }
            \hline
            \emph{Symbol} & \emph{Description} & \emph{Units} \\ \hline
            $c_{\!p}$ & Specific heat at constant pressure & J kg\textsuperscript{-1} K\textsuperscript{-1} \\ \hline
            $h$ & Specific enthalpy & J kg\textsuperscript{-1} \\ \hline
            $\kappa$ & Thermal conductivity & W m\textsuperscript{-1} K\textsuperscript{-1} \\ \hline
            $m$ & Mass & kg \\ \hline
            $\dot m$ & Mass flow rate & kg s\textsuperscript{-1} \\ \hline
            $n$ & Number of elements (e.g. PCM capsules) &  \\ \hline
            $r$ & Internal radius & m \\ \hline
            $\rho$ & Density & kg m\textsuperscript{-3} \\ \hline
            $t$ & Time & s \\ \hline
            $P$ & Pressure & Pa \\ \hline
            $\dot Q$ & Transferred cooling power & W \\ \hline
            $R$ & Thermal resistance & K W\textsuperscript{-1}  \\ \hline
            $T$ & Temperature & ºC  \\ \hline
            $U$ & Internal energy & J  \\ \hline
            $V$ & Volume & m\textsuperscript{3} \\ \hline
    \end{tabu}}
\end{table}

\begin{table}[H]
    \centering
    \caption{Subscript/superscript notation}
    \label{tabSubscriptsSuperscripts}
    \scalebox{0.7}[0.7]{\tabulinesep=0.5mm
        \begin{tabu} { L{1.3cm}  L{6cm} | L{1.3cm}  L{6cm} }
            \hline
            \multicolumn{2}{c}{\emph{\bf Subscripts}} & \multicolumn{2}{c}{\emph{\bf Superscripts}} \\ \hline
            \emph{Symbol} & \emph{Description} & \emph{Symbol} & \emph{Description}\\ \hline
            $\tinyINT$ & Intermediate fluid & $\tinyIN$ & Input/Inlet \\ \hline
            $\tinyREF$ & Refrigerant & $\tinyOUT$ & Output/Outlet \\ \hline
            $\tinyREFVAP$ & Refrigerant in vapour phase & $\tinyCOND$ & Conduction \\ \hline
            $\tinyREFTWO$ & Refrigerant in two phase & $\tinyCONV$ & Convection \\ \hline
            $\tinySEC$ & Secondary fluid & $\tinyINT$ & Internal \\ \hline
            $\tinyPCM$ & PCM & $\tinyEXT$ & External \\ \hline
            $\tinyPCMLIQ$ & PCM in liquid phase & $\tinyLAT$ & Latent state \\ \hline
            $\tinyPCMSOL$ & PCM in solid phase & $\tinyLATMAX$ & Maximum enthalpy latency point \\ \hline
            $\tinyLAY$ & Layer of PCM-capsule volume & $\tinyLATMIN$ & Minimum enthalpy latency point \\ \hline
            & & $\tinyMAX$ & Maximum \\ \hline
            & & $\tinyMIN$ & Minimum \\ \hline
            & & $\tinyWALL$ & Wall \\ \hline
    \end{tabu}}
\end{table}

\subsection{Design specifications of the TES system}\label{subsecDesignSpecifications}

The PCM-based TES tank under study is the main component in the upgrading process undertaken on an experimental refrigeration facility, located at the Department of Systems Engineering and Automation at the University of Seville. A detailed description of the pre-existing plant can be found in previous works by Bejarano \emph{et al.} \cite{GB_JE_2015,bejarano2015multivariable}. Additionally, the technical details of the embedding of the TES tank in the plant can be found in a previous work by the authors \cite{Bejarano2017NovelSchemePCM}. One of the particularly relevant aspects described in the latter is that the TES tank has been projected to work in parallel with the existing refrigerated chamber. This fact, among many others, dictated the design specifications of the TES system. Specifically, the refrigeration system should satisfy cooling demand around $-20^{\circ}\mathrm{C}$, in accordance with typical reference levels for refrigeration systems.

\begin{figure}[h]
    \centerline{\includegraphics[width=7cm,trim = 0 0 0 0,clip]
    {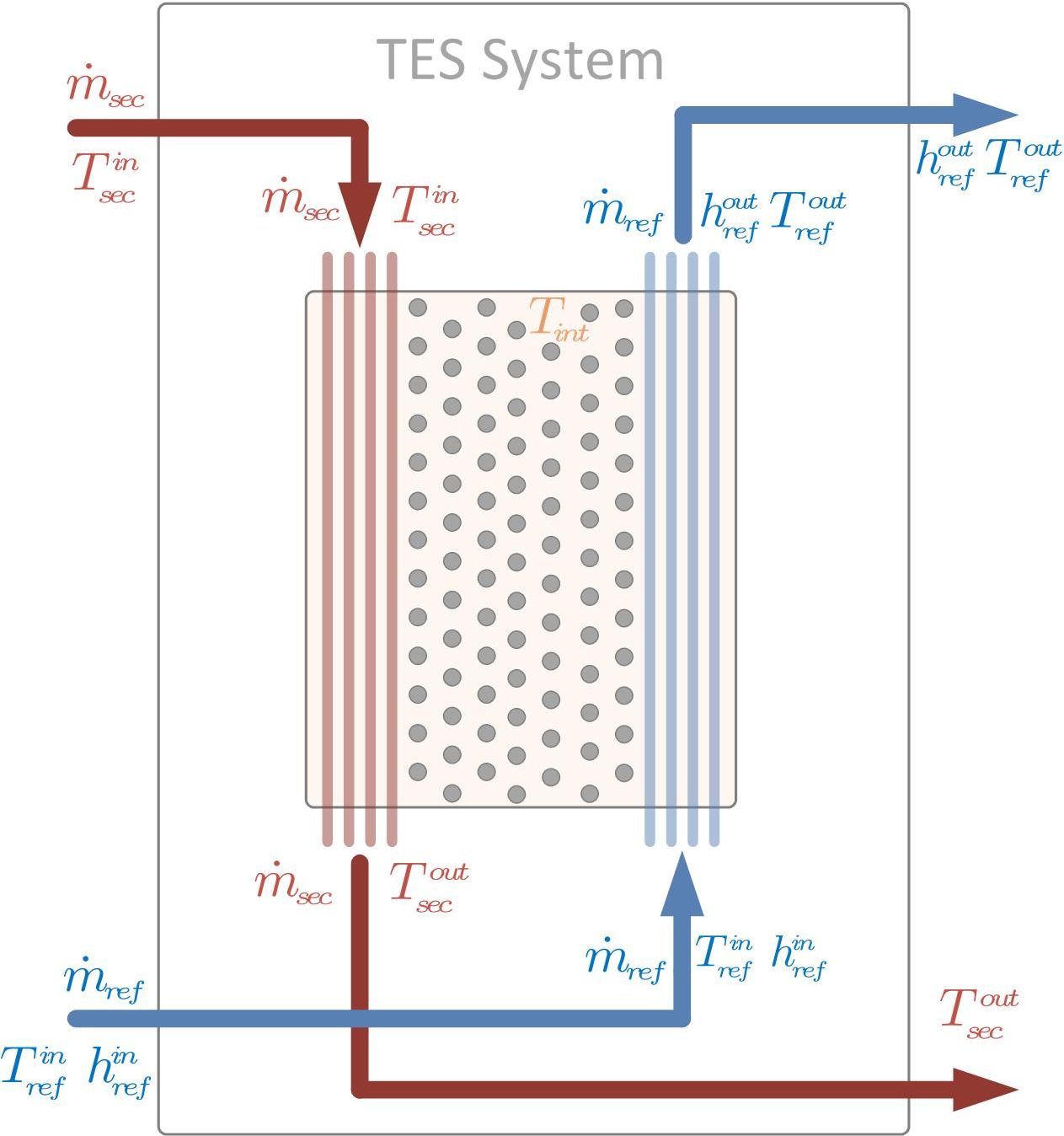}}
    \caption{Schematic picture of the proposed configuration of the TES tank and input-output conceptualization of the TES system.}
    \label{figTEStank}
\end{figure}

There are several alternatives regarding the selection of the phase-change material \cite{oro2012review}: bulk PCM inside the tank; macro-encapsulated PCM, where the raw material is enclosed in some form of package such as tubes, pouches, spheres, panels or other kind of receptacle, usually larger than 1~cm in diameter; finally, micro-encapsulated PCM, similar to the previous one, but where the encapsulation diameter is in the order of millimetres. The macro-encapsulated PCM is the most common form of encapsulation, whose major advantages are the ease of manufacturing and marketing, as well as the flexibility on the design of the PCM nodule shape. In our design, the TES tank is filled with macro-encapsulated efficient PCM, where the raw material is enclosed in spherical polymer balls. The PCM spheres are bathed in the so-called intermediate fluid, featuring high thermal conductivity and low heat capacity. In addition to the PCM balls, two bundles of pipes, corresponding to the refrigerant and secondary fluid, are also in contact with the intermediate fluid and running through the tank. This way, when the refrigerant is flowing, the tank acts as an evaporator, where the refrigerant evaporates while extracting heat from the intermediate fluid (and eventually from the PCM capsules). On the other hand, when the secondary fluid is flowing, it transfers heat to the intermediate fluid and the PCM spheres. Figure \ref{figTEStank} describes the schematic setup of the TES system\footnote{Please, notice that this is just a conceptual scheme. Of course, from a constructive point of view, all pipes must be distributed in such a way that homogeneous heat transfer is attained all around the TES tank.}.

Conceptual input-output representation of the system can also be noticed in Figure \ref{figTEStank}. The inputs are the refrigerant mass flow rate, $\dot m_\tinyREF(t)$, and secondary fluid mass flow rate, $\dot m_\tinySEC(t)$, essentially used as control inputs during the charging or discharging cycle, respectively. Other input signals are the inlet temperature of the secondary fluid, $T_\tinySEC^\tinyIN(t)$, the inlet temperature and the specific enthalpy of the refrigerant, $T_\tinyREF^\tinyIN(t)$ and $h_\tinyREF^\tinyIN(t)$. Clearly, the different fluid pressures also affect the system, but they are not shown in Figure \ref{figTEStank}, since they are assumed to be constant, according to the equilibrium conditions the whole TES-backed-up refrigeration plant is operating at. The relevant output signals are also shown in Figure \ref{figTEStank}, namely, the outlet thermodynamic conditions of both the refrigerant and the secondary fluid can be mentioned.

\begin{figure}[htbp]
    \centerline{\includegraphics[width=7cm]
    {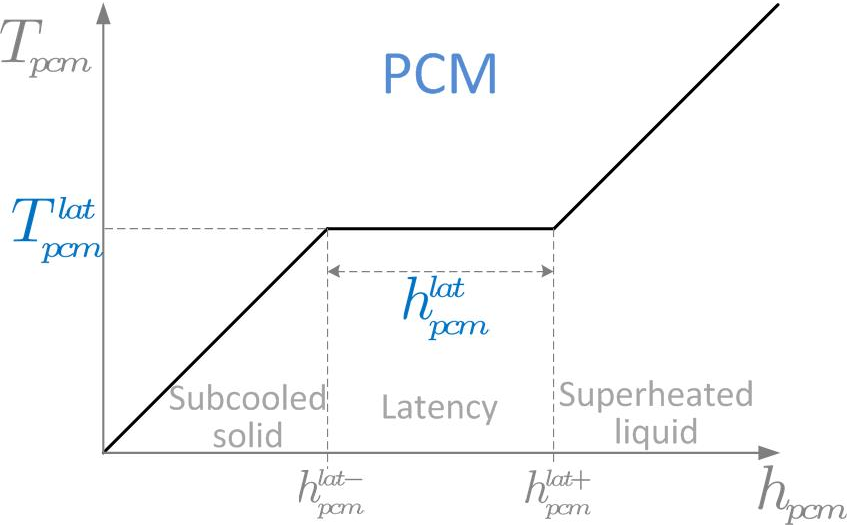}}
    \caption{Temperature-enthalpy diagram of the PCM.}
    \label{figThDiagrams}
\end{figure}

During normal operation of our system, both materials, the refrigerant and the PCM, may experience phase change. In particular, the PCM may be in solid phase, liquid phase, or in transition between both. Figure \ref{figThDiagrams} shows the temperature-enthalpy diagram of the PCM. On the other hand, the refrigerant may appear in liquid phase, vapour phase, or two-phase.

The values of the main variables at the nominal operating point of the experimental plant are shown in Table \ref{tabNominalOperatingConditions}.

\begin{table}[h]
    \centering
    \caption{Nominal operating conditions}
    \label{tabNominalOperatingConditions}
    \scalebox{0.7}[0.7]{ \tabulinesep=0.5mm{}
    \begin{tabu} {L{1.5cm}  L{11cm} L{1.5cm} L{2cm}}
        \emph{Symbol} & \emph{Description} & \emph{Value} & \emph{Units} \\
        \Xhline{5\arrayrulewidth}
        $\dot m_\tinyREF$ & Refrigerant mass flow rate & 0.00918 & kg   s\textsuperscript{-1} \\
        \Xhline{2\arrayrulewidth}
        $h_\tinyREF^\tinyIN$ & Specific enthalpy of the refrigerant at the TES tank inlet & 255000 & J kg\textsuperscript{-1} K\textsuperscript{-1} \\
        \Xhline{2\arrayrulewidth}
        $P_\tinyREF$ & Refrigerant pressure & 126500 & Pa \\
        \Xhline{2\arrayrulewidth}
        $T_\tinyREF^\tinyIN$ & Temperature of the refrigerant at the TES tank inlet & -41.08 & ºC \\
        \Xhline{3\arrayrulewidth}
        $\dot m_\tinySEC$ & Secondary mass flow rate & 0.074 & kg s\textsuperscript{-1} \\
        \Xhline{2\arrayrulewidth}
        $P_\tinySEC$ & Secondary fluid pressure & 100000 & Pa \\
        \Xhline{2\arrayrulewidth}
        $T_\tinySEC^\tinyIN$ & Temperature of the secondary fluid at the TES tank inlet & -20 & ºC \\
        \Xhline{2\arrayrulewidth}
    \end{tabu}}
\end{table}

The TES system has been projected following a detailed set of specifications. In particular the PCM capsules are filled with an ad-hoc designed vegetable oil enclosed in spherical polymer capsules, made of high-density polyethylene. The chosen vegetable oil is a phase change material designed taking the \emph{PureTemp-37}$^{\tiny{\textregistered}}$ \cite{PureTemp-37} as a reference, with the same properties except for the melting point, which had to be adapted to our application, being -30ºC instead of -37ºC, due to the features of the pre-existing refrigeration plant and the fluid acting as refrigerant, as well as the temperature at which the cooling power is intended to be provided (around -20ºC). Please notice that this work describes a preliminary design, very well suited to simulations tasks and to solve simulation issues, given that, at the present time, the TES tank has not been built yet. 

The fluid chosen as refrigerant is R404a, which is a fluid commonly used in refrigeration cycles, specifically in medium-low temperature facilities. The intermediate fluid is a 60\% v/v ethylene glycol aqueous solution with very high thermal conductivity, providing efficient heat transfer between the refrigerant and the PCM capsules during the charging cycle, or between the PCM capsules and the secondary fluid during the discharging cycle. The mixture has low freezing temperature which meets the requirements inside the tank. The secondary fluid is a 60\% v/v propylene glycol aqueous solution. The pipes are all made of carbon steel.

The PCM properties and design parameters considered within the simulations are shown in Tables \ref{tabPCMProperties} and \ref{tabDesignParameters}.

\begin{table}[htbp]
    \centering
    \caption{Phase change material properties}
    \label{tabPCMProperties}
    \scalebox{0.6}[0.6]{ \tabulinesep=0.5mm{}
        \begin{tabu} {L{1.5cm} L{12.5cm} L{2cm} L{2cm}}
            \emph{Symbol} & \emph{Description} & \emph{Value} & \emph{Units} \\ \Xhline{3\arrayrulewidth}
            $c_{\!p_{\,\tinyPCMLIQ}}$ & Specific heat of the PCM in liquid phase at constant pressure & 1990 & J kg\textsuperscript{-1} K\textsuperscript{-1} \\
            \Xhline{2\arrayrulewidth}
            $c_{\!p_{\,\tinyPCMSOL}}$ & Specific heat of the PCM in solid phase at constant pressure & 1390 & J kg\textsuperscript{-1} K\textsuperscript{-1} \\
            \Xhline{2\arrayrulewidth}
            $h_\tinyPCM^\tinyLAT$ & Specific enthalpy of fusion (latent phase) of the PCM & 145000 & J kg\textsuperscript{-1} \\
            \Xhline{2\arrayrulewidth}
            $T_\tinyPCM^\tinyLAT$ & Phase change temperature & -30 & ºC\\
            \Xhline{2\arrayrulewidth}$\kappa_\tinyPCM^\tinyLATMAX$ &  Thermal conductivity of PCM in liquid phase & 0.15 & W m\textsuperscript{-1} K\textsuperscript{-1} \\
            \Xhline{2\arrayrulewidth}
            $\kappa_\tinyPCM^\tinyLATMIN$ &  Thermal conductivity of PCM in solid phase & 0.25 &  W m\textsuperscript{-1} K\textsuperscript{-1} \\
            \Xhline{2\arrayrulewidth}
            $\rho_\tinyPCM^\tinyLATMAX$ & Density of the PCM in liquid phase & 880 & kg m\textsuperscript{-3}\\
            \Xhline{2\arrayrulewidth}
            $\rho_\tinyPCM^\tinyLATMIN$ & Density of the PCM in solid phase & 970 & kg m\textsuperscript{-3}\\
            \Xhline{2\arrayrulewidth}
    \end{tabu}}
\end{table}

\begin{table} [htbp]
    \centering
    \caption{Design parameters}
    \label{tabDesignParameters}
    \scalebox{0.6}[0.6]{ \tabulinesep=0.5mm {}
    \begin{tabu} {L{1.5cm} L{14cm} L{3.5cm} L{2cm}}
    \emph{Symbol} & \emph{Description} & \emph{Value} & \emph{Units} \\ \Xhline{3\arrayrulewidth}
    $h_\tinyREF^\tinyLAT$ & Specific enthalpy of vaporization (latent phase) of the refrigerant  & 197770 & J kg\textsuperscript{-1} \\ \Xhline{2\arrayrulewidth}
    $m_\tinyINT$ & Mass of the intermediate fluid & 56.37 & kg \\ \Xhline{2\arrayrulewidth}
    $n_\tinyPCM$ & Number of PCM capsules & 400 & \\ \Xhline{2\arrayrulewidth}
    $n_\tinyREF$ & Number of refrigerant pipes & 50 & \\ \Xhline{2\arrayrulewidth}
    $n_\tinySEC$ & Number of secondary fluid pipes & 50 & \\ \Xhline{2\arrayrulewidth}
    $r_{\!\tinyPCM}^{\,\tinyMAX}$ & Maximum internal radius of the PCM capsules & 0.0285 & m \\    \Xhline{2\arrayrulewidth}
    $r_{\!\tinyPCM}^{\,\tinyMIN}$ & Minimum internal radius of the PCM capsules & 0.02759 & m \\    \Xhline{2\arrayrulewidth}
    \end{tabu}}
\end{table}


\section{Discrete model of the TES system} \label{secDiscreteModel}

The discrete model to be used is built upon the continuous model of the TES system that was thoroughly described in the aforementioned work by Bejarano \emph{et al.} \cite{Bejarano2017NovelSchemePCM}. This continuous model was based on other works found in the related literature, where the freezing and melting processes of PCM capsules were studied and experimentally validated
\cite{bedecarrats1996phase,bedecarrats2009bstudy,ismail2009numerical,temirel2017solidification,amin2014effective}.

The main limitation of this continuous model was evinced in the mentioned work by Bejarano {\emph et al.} \cite{Bejarano2017NovelSchemePCM}, that is to say, its applicability is limited to full charging or discharging operations. A discrete adaptation of the model was also developed, in this case with general applicability. Both models were analysed and compared in the mentioned work \cite{Bejarano2017NovelSchemePCM}.

In the continuous model, the behaviour of the PCM capsules is simplified assuming that only one single inward freezing/melting front is present, being only valid for modelling a continuous charging/discharging cycle, starting from a discharged/charged steady state, as already mentioned.

The discrete model to be summarised here is not restricted to full charging or discharging operations, but it can also represent the dynamic evolution of the system during any series of partial charging/discharging operations, in such a way that an arbitrary number of moving freezing/melting boundaries could be present at the same time inside the PCM capsules.

The intermediate fluid is assumed to be completely still inside the TES tank, at constant atmospheric pressure, though closed, bathing the PCM capsules and the refrigerant and secondary fluid bundles of tubes. Its high thermal conductivity allows us to assume that it has a homogeneous temperature around the whole tank volume.

The energy-balance equation of the intermediate fluid is shown in Equation
\eqref{eqPrincipalDobleDiscreta}, where the main factors are detailed below.

\begin{equation}
   T_\tinyINT(t+\Delta t) = T_\tinyINT(t) - \frac{\Delta t}{m_\tinyINT c_{\!p_\tinyINT}}\,
   \Big [ n_\tinyREF \dot Q_\tinyREF(t) + n_\tinySEC \dot Q_\tinySEC(t) + n_\tinyPCM \dot Q_\tinyPCM(t) \Big ]
   \label{eqPrincipalDobleDiscreta}
\end{equation}

\begin{itemize}
   \item $T_\tinyINT(t)$: Temperature of the intermediate fluid (assumed to be instantaneously homogeneous inside the intermediate fluid reservoir).
   \item $\dot Q_\tinyREF(t)$: Cooling power transferred from each single refrigerant pipe to the intermediate fluid (typically positive).
   \item $\dot Q_\tinySEC(t)$: Cooling power transferred from each single secondary fluid pipe to the intermediate fluid (typically negative).
   \item $\dot Q_\tinyPCM(t)$: Cooling power transferred from each PCM capsule to the intermediate fluid (positive during discharging cycle, negative during charging cycle).
\end{itemize}

During the charging cycle, all secondary mass flow is stopped ($\dot Q_\tinySEC(t)=0$), the refrigerant is providing cooling power to the intermediate fluid ($\dot Q_\tinyREF(t)>0$), while each one of the PCM capsules is contributing to store this cold energy, taking it from the intermediate fluid ($\dot Q_\tinyPCM<0$). On the other hand, during the discharging cycle, the refrigerant mass flow is stopped ($\dot Q_\tinyREF(t)=0$), the PCM capsules transfer cold energy to the intermediate fluid ($\dot Q_\tinyPCM(t)>0$), while the secondary fluid is taking cooling power from the intermediate fluid ($\dot Q_\tinySEC(t)<0$).

The discrete model uses a sampling interval $\Delta t = 2$~s, which can be viewed as a suitable trade-off between simulation accuracy and expected overall simulation time of the whole system. In fact, regarding the accuracy in the heat transfer dynamics at the PCM-nodules level, $\Delta t = 2$~s has been found to be about the maximum acceptable time interval. A shorter simulation interval would lead, of course, to more precise simulations, but at the expense of an excessively high computational cost at the top simulation level.

\subsection{PCM-intermediate fluid dynamics}

\begin{figure}[htbp]
    \centerline{\includegraphics[width=10cm,trim={0 115 0 0},clip]
    {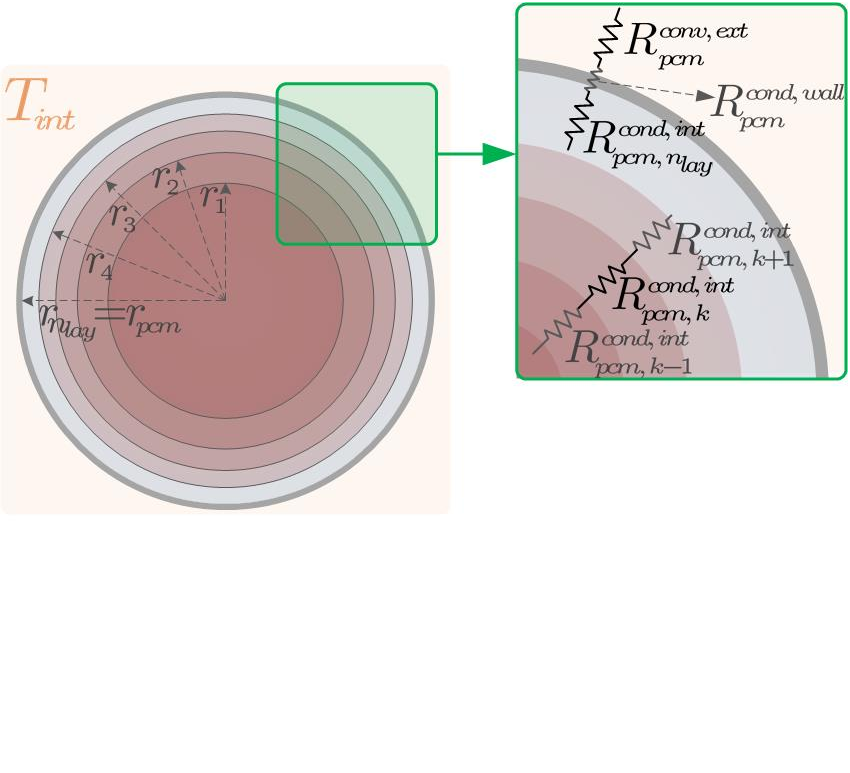}}
    \caption{Scheme of the discretised PCM capsule for $n_\tinyLAY = 5$, with a detail of the involved thermal resistances.}
    \label{figPCMcapsuleOnion}
\end{figure}

This discrete model is also a volume-discrete one, where a discretisation of the continuous PCM-capsule volume is defined, according to a predefined number of layers, $n_\tinyLAY$, each one of them including the exact same PCM mass $m_\tinyLAY$. The discretised PCM capsule is represented in Figure \ref{figPCMcapsuleOnion} for $n_\tinyLAY=5$, showing also a detail of the involved thermal resistances.

The system state of this discrete model is represented by the specific enthalpy of each layer, $h_\tinyPCMK(t)$, together with the temperature of the intermediate fluid, $T_\tinyINT(t)$.

\vspace{1em} 

Energy balance in each particular layer $k$ is ruled by Equation Set \eqref{eq:hdot},
for $k=1, 2, \,\ldots ,\,n_\tinyLAY$.

\begin{equation}
   \begin{aligned}
   		h_\tinyPCMK(t+\Delta t) \,&=\, h_\tinyPCMK(t) + \frac{1}{\rho_\tinyPCMK(t)\, V_\tinyPCMK(t)} \left( \dot Q_\tinyPCMK^{\,\tinyEXT}(t) - \dot Q_\tinyPCMK^{\,\tinyINT}(t) \right) \Delta t \\
   		\dot Q_\tinyPCMK^{\,\tinyEXT}(t) \,&=\, \frac{T_\tinyPCMKMas(t) - T_\tinyPCMK(t)}{R_\tinyPCMKMas^\tinyCOND(t)} \\
   		\dot Q_\tinyPCMK^{\,\tinyINT}(t) \,&=\, \frac{T_\tinyPCMK(t) - T_\tinyPCMKMenos(t)}{R_\tinyPCMK^\tinyCOND(t)} \\
   \end{aligned}
   \label{eq:hdot}
\end{equation}

In Equation Set \eqref{eq:hdot} $V_\tinyPCMK (t)$ is the volume of every individual layer ($V_\tinyPCMK(t) = \frac{m_\tinyLAY}{\rho_\tinyPCMK(t)}$), while $\dot Q_\tinyPCMK^{\,\tinyEXT}(t)$ is the cooling power transferred to the neighbour outer layer (layer number $k\!+\!1$) and $\dot Q_\tinyPCMK^{\,\tinyINT}(t)$ is the cooling power received from the inner neighbour layer (layer number $k\!-\!1$). $R_\tinyPCMK^\tinyCOND(t)$ represents the thermal resistance linked to the heat conduction properties of layer $k$ of the PCM ball.

The inner-most layer verifies $\dot Q_\tinyPCMOne^{\,\tinyINT}(t)\!=\! 0$, while the outermost layer is that one ultimately exchanging cooling power with the intermediate fluid, in such a way that $\dot Q_\tinyPCMNLAY^{\,\tinyEXT}(t)\!=\! \dot Q_\tinyPCM(t)$. A particular expression is needed for $\dot Q_\tinyPCMNLAY^{\,\tinyEXT}(t)$ as shown by Equation \eqref{eq:Q_pcm_nlay}.

\begin{equation}
   \dot Q_\tinyPCMNLAY^{\,\tinyEXT}(t)\!=\!
    \frac{T_\tinyINT(t) - T_\tinyPCMNLAY(t)}{R_\tinyPCM^\tinyCONDWALL + R_\tinyPCM^\tinyCONVEXT(t)}
    \label{eq:Q_pcm_nlay}
\end{equation}

The denominator represents the global thermal resistance in the heat transfer process between the outermost layer of the PCM nodule and the intermediate fluid. Two terms are involved, namely, a conduction term, given by the thickness and thermal conductivity of the nodule polymer coating, followed by an external, natural convective term, associated with the intermediate fluid itself.

\vspace{2em}

For each layer $k$, the dependence between its temperature and its specific enthalpy
has three alternatives, according to Figure \ref{figThDiagrams}, as indicated in Equation Set \eqref{eq:h_T}.

\begin{equation}
   \begin{aligned}
        T_\tinyPCMK(t) &= T_\tinyPCM^\tinyLAT & \qquad & \text{if} \quad &h_\tinyPCM^\tinyLATMIN \le &h_\tinyPCMK(t) \le &h_\tinyPCM^\tinyLATMAX \\
        T_\tinyPCMK(t) &= T_\tinyPCM^\tinyLAT - \frac{h_\tinyPCM^\tinyLATMIN - h_\tinyPCMK(t)}{c_{\!p_\tinyPCMSOL}} & \qquad & \text{if} \quad  &h_\tinyPCM^\tinyLATMIN > & h_\tinyPCMK(t) & \\
        T_\tinyPCMK(t) &= T_\tinyPCM^\tinyLAT + \frac{h_\tinyPCMK(t) - h_\tinyPCM^\tinyLATMAX}{c_{\!p_\tinyPCMLIQ}} & \qquad & \text{if} \quad  & & h_\tinyPCMK(t) > &h_\tinyPCM^\tinyLATMAX
    \end{aligned}
    \label{eq:h_T}
\end{equation}

The detailed analysis and comparison of the continuous and discrete models can be found in the previously referred work by Bejarano \emph{et al.} \cite{Bejarano2017NovelSchemePCM}.

\subsection{Refrigerant - intermediate fluid dynamics}

The \emph{moving boundary} approach \cite{Rasmussen2005,Liang2010} is applied to model the refrigerant behaviour along its way through the tank. By design, the refrigerant enters the tank as two-phase fluid and it usually comes out as slightly superheated vapour, but, alternatively, it may also come out as two-phase fluid.

According to this configuration, expressions describing heat transfer between the refrigerant and the intermediate fluid need to be provided for both zones, the superheated vapour zone and the two-phase zone, since the global cooling power transferred is computed as shown in Equation \eqref{eq:QREF}.

\begin{equation}
    \dot Q_\tinyREF(t) \,=\, \dot Q_\tinyREFTWO(t) + \dot Q_\tinyREFVAP(t)
    \label{eq:QREF}
\end{equation}

In the case of the two-phase zone, the involved fluids, i.e. the intermediate fluid and the refrigerant in latent phase, is each one at instantaneously homogeneous temperature along the heat transfer area; hence, mere integration of the general heat transfer expression suffices, as described in Equation \eqref{eq:QREFTWO}.

\begin{equation}
    \dot Q_\tinyREFTWO(t) \,=\, \frac{T_\tinyINT(t) - T_\tinyREF^\tinyIN(t)}{R_\tinyREFTWO^\tinyCONVINT\!(t)  + R_\tinyREFTWO^\tinyCONDWALL\!(t)  + R_\tinyREFTWO^\tinyCONVEXT\!(t)}
    \label{eq:QREFTWO}
\end{equation}

The denominator represents the global thermal resistance in the heat transfer process between the refrigerant and intermediate fluid, where three thermal resistances are arranged in series. There is a conduction term, given by the geometry and thermal conductivity of the refrigerant pipe wall, together with internal and external convective terms (forced and natural, respectively), which depend on the length of the refrigerant two-phase zone and the corresponding convective heat transfer coefficients.

An additional relation is required in this case, due to the refrigerant moving boundary, which is described by Equation \eqref{eq:QREFTWO_add}.

\begin{equation}
    \dot Q_\tinyREFTWO(t) \,=\, \dot m_\tinyREF\!(t)\, (h_\tinyREF^\tinyLATMAX - h_\tinyREF^\tinyIN(t))
    \label{eq:QREFTWO_add}
\end{equation}

On the other hand, the \emph{effectiveness-NTU} method is used to describe heat transfer at the superheated vapour zone, where the intermediate fluid acts as the hot source, while the refrigerant is the cold source. As mentioned before, the intermediate fluid is completely still and considered, by comparison, a very large mass with instantaneously homogeneous temperature. According to this, the corresponding transferred cooling power is estimated according to Equation \eqref{eq:eNTU_epsilon_REFVAP}.

\begin{equation}
   \dot Q_\tinyREFVAP(t) \,=\, \left(1-\text{e}^{\frac{-1}{R_\tinyREFVAP(t)\,c_{\!p_{\,\tinyREFVAP}}\,\dot m_\tinyREF(t)}} \right ) \, c_{\!p_{\,\tinyREFVAP}}\,\dot m_\tinyREF(t)\,(T_\tinyINT(t) - T_\tinyREF^\tinyIN(t))
   \label{eq:eNTU_epsilon_REFVAP}
 \end{equation}

$R_\tinyREFVAP(t)$ represents the global thermal resistance in the heat transfer process between the refrigerant and the intermediate fluid, at the superheated vapour zone. The three terms involved are indicated in Equation \eqref{eq:R_REFVAP}.

\begin{equation}
    R_\tinyREFVAP\!(t) \,=\, R_\tinyREFVAP^\tinyCONVINT\!(t)  + R_\tinyREFVAP^\tinyCONDWALL\!(t)  + R_\tinyREFVAP^\tinyCONVEXT\!(t)
    \label{eq:R_REFVAP}
\end{equation}

\subsection{Secondary fluid - intermediate fluid dynamics}

During the discharging cycle, the secondary fluid is always in liquid phase, flowing through a bundle of tubes, at constant pressure, submerged in the still intermediate fluid. The \emph{effectiveness-NTU} method is used again, as shown in Equation \eqref{eq:eNTU_epsilon_SEC}, where the secondary fluid acts as the hot source, while the intermediate fluid is the cold source.

\begin{equation}
    \dot Q_\tinySEC(t) \,=\, \left(1-\text{e}^{\frac{-1}{R_\tinySEC(t)\,c_{p_{\,\tinySEC}}\,\dot m_\tinySEC(t)}} \right) \, c_{p_{\,\tinySEC}}\,\dot m_\tinySEC(t)\,(T_\tinyINT(t) -  T_\tinySEC^\tinyIN(t)) \\
    \label{eq:eNTU_epsilon_SEC}
\end{equation}

$R_\tinySEC(t)$ represents the global thermal resistance in the heat transfer between the secondary and intermediate fluids, where once again the three terms shown in Equation \eqref{eq:R_SEC} are involved.

\begin{equation}
    R_\tinySEC(t) \,=\, R_\tinySEC^\tinyCONVINT(t)  + R_\tinySEC^\tinyCONDWALL  + R_\tinySEC^\tinyCONVEXT(t)
    \label{eq:R_SEC}
\end{equation}

$\dot Q_\tinySEC(t)$ is also related to the inlet-outlet temperature difference in the secondary fluid pipes, through the heat capacity rate, as described in Equation \eqref{eq:Q_SEC_aux}, or alternatively through the mass average temperature inside the pipe and the pipe wall temperature, as indicated in Equation \eqref{eq:Q_SEC_aux2}.

\begin{equation}
    \dot Q_\tinySEC(t) \,=\, c_{p_{\,\tinySEC}}\,\dot m_\tinySEC(t)\,(T_\tinySEC^\tinyOUT(t) - T_\tinySEC^\tinyIN(t))
    \label{eq:Q_SEC_aux}
\end{equation}

\begin{equation}
    \dot Q_\tinySEC(t) \,=\, \frac{T_\tinySEC^\tinyWALL(t) - \frac{1}{2}\left(T_\tinySEC^\tinyIN(t) + T_\tinySEC^\tinyOUT(t)\right)}{R_\tinySEC^\tinyCONVINT(t) + R_\tinySEC^\tinyCONDWALL}
    \label{eq:Q_SEC_aux2}
\end{equation}


\section{Time-efficient discrete model of the TES system} \label{secEfficientDiscreteModels}

Notwithstanding the benefits of allowing a more complete description of the system behaviour, the computational cost of the discrete model is too high for the intended purpose. In particular, keeping in mind that the modelled TES system is projected just as a component part of a more complex refrigeration system, for which efficient energy management strategies are expected to be implemented in the near future. As mentioned in Section \ref{Introduction}, the time scale of such management strategies may be in the range of weeks or months; it implies that small time steps around few seconds, which are needed by the discrete model (following a fixed-step integration procedure) to properly describe all dynamics involved, including the layers in sensible zone, are inconceivable. This amply justifies the need for an "accelerated" version of the previous discrete model, where the fast dynamics may be disregarded and the time step can be extended to the order of minutes.

The same layered structure of the PCM nodules will be kept, but a much loose-fitting time discretisation is being applied, besides some other approximations. The discrete model described in Section \ref{secDiscreteModel} includes an energy balance for every layer, shown in Equation Set \eqref{eq:hdot}, which is the same for the layers in latent zone and those in sensible zone. The dynamics related to the layers in latent zone turn out to be slower than those of the layers in sensible zone, due to the continuous change in their temperature, which affects the calculation of the cooling powers transferred, not only between them but also at the moving freezing/melting interface. As stated in Section \ref{secDiscreteModel}, the difference equations in the discrete domain are integrated using a fixed-step procedure, with a sampling interval of 2 s. This is the maximum time step required to properly integrate the difference equations, which depends on the number of layers, being smaller as the latter increases. Variable-step integration methods could be applied, but in any case the sampling time would be adapted to the fastest system dynamics.

The idea of the proposed time-efficient discrete model starts from the premise that it is necessary to neglect the fastest dynamics to achieve longer time intervals and accelerate the simulation of the discrete model. Therefore, thermal dynamics of the layers in sensible zone are considered as negligible, imposing instead a steady-state conduction model. Furthermore, regarding the layers in sensible zone, it is observed that all consecutive layers can be clustered into a single thermal-conduction layer, whose thermal resistance does not vary until the first inward layer in latent zone drains away all latent energy. Being one of the thermal focuses constant (that related to the phase change temperature), the evolution of the cooling power transferred between the PCM capsule and the intermediate fluid is only due to the slow evolution of the temperature of the intermediate fluid, provided that the mentioned first inward layer remains in latent zone. Therefore, it can be assumed that cold energy is transferred at a constant rate during a certain time interval, when the set of layers in sensible zone does not vary. The duration of this time interval can be estimated considering the calculated cooling power and the remaining latent energy of the first inward layer in latent zone, thus the integration of the difference equations of the model within this time interval can be performed in one step. It definitely allows to accelerate the simulation and consider longer time periods, which may be subdivided into as many intervals as necessary to describe the transition from latent to sensible zone of every layer. The proposed method can be seen as an adaptive variable-step integration procedure, but the main difference with respect to conventional variable-step integration methods lies in the fact that the time intervals are predicted according to the system state, not being calculated based on errors corresponding to a previous integration. Moreover, the proposed procedure does not involve recalculation, as conventional variable-step methods do.

Therefore, the two main assumptions made to develop the time-efficient discrete model are summarised in the following:

\begin{itemize}

    \item The dynamics of the layers in sensible zone are considered as negligible, calculating their temperature according to a steady-state conduction model.

    \item Cold energy is assumed to be transferred at constant rate during a certain time interval, while the set of layers in latent and sensible zone does not vary.

\end{itemize}

It will be shown in Section \ref{secComputationalCost} that these assumptions allow the simulation time to be notably reduced, whereas the overall accuracy when describing the system behaviour is retained. Time periods are now increased to tens or hundreds times the initial one used in the original discrete model; simulations using $\Delta t =\{60 \, \text{s},\,180\, \text{s},\,300\, \text{s}\}$ will be reported in Section \ref{secAccuracyAssessment}. In particular, a value such $\Delta t = 300\, \text{s}$ is just below the acceptable sampling limit imposed by the charge/discharge dynamics inherent to the PCM nodules and also defined by the nominal operating conditions shown in Table \ref{tabNominalOperatingConditions}, namely the inlet conditions of the refrigerant (charge) and the secondary fluid (discharge).  The difference between the inlet temperature of the corresponding fluid and the melting temperature of the PCM mainly defines the charging/discharging speed, which limits the maximum sampling time to be considered.

The description of the recursive algorithm implementing this "accelerated" discrete model will be given in a step-by-step sketch, as follows:

\begin{enumerate}

    \item
        Starting from a given enthalpic state of the layered PCM capsule, an inward scanning sequence of the layers is performed, looking for the first layer $k_0$ in latent zone, as indicated in Equation \eqref{eq:k0}.

    \begin{equation}
        k_0 \,=\, \operatorname{max}\left\{\, k \in \{1,\ldots n_\tinyLAY\} \;\;\mid\;\;
        h_\tinyPCM^\tinyLATMIN \le h_\tinyPCMK(t) \le h_\tinyPCM^\tinyLATMAX \,\right\}
        \label{eq:k0}
    \end{equation}

    \item
    	Thermal resistance $R_\tinyPCMKCero^\tinyCOND$ is computed, corresponding to the spherical shell comprising layers outside layer $k_0$: $k \in \{k_0\!+\!1,\ldots n_\tinyLAY\}$, taken as a single, clustered thermal-conduction layer.

    \item
    	The cooling power exchanged between the intermediate fluid and $k_0$ layer is estimated as shown in Equation \eqref{eq:Q_pcm_k0}.

    \begin{equation}
        \dot Q_\tinyPCMKCero(t) \,=\, \frac{T_\tinyINT(t) - T_\tinyPCMKCero(t)}{R_\tinyPCMKCero^\tinyCOND(t) + R_\tinyPCM^\tinyCONDWALL + R_\tinyPCM^\tinyCONVEXT(t)}
        \label{eq:Q_pcm_k0}
    \end{equation}

    \item
    	It is assumed that, during the whole period $\Delta t$, layer $k_0$ will remain in latent zone, in such a way that cold energy is transferred at the constant rate $\dot Q_\tinyPCMKCero(t)$, as indicated in Equation \eqref{eq:U_pcm_k0}.

    \begin{equation}
        \Delta U_\tinyPCMKCero(t) \,\approx\, \dot Q_\tinyPCMKCero(t) \, \Delta t \label{eq:U_pcm_k0}
    \end{equation}

		This is the main approximation assumed by the time-efficient discrete model, since actually the cooling power varies continuously. It will be shown below that the relative errors generated by this approximation are small enough to be assumed. This would be, eventually, the essential source of error of the "accelerated" model with respect to the original discrete one.

    \item
    	The specific enthalpy of layer $k_0$ is updated accordingly, as shown in Equation \eqref{eq:h_pcm_k0_forward}.

    \begin{equation}
        h_\tinyPCMKCero(t\!+\!\Delta t) \,=\,  h_\tinyPCMKCero(t) + \Delta U_\tinyPCMKCero(t)\, \frac{1}{\rho_\tinyPCMKCero(t)\, V_\tinyPCMKCero(t)}
        \label{eq:h_pcm_k0_forward}
    \end{equation}

    \item
    	At this point, two possibilities arise:

    \begin{itemize}

        \item
       		Layer $k_0$ remains in latent zone: $h_\tinyPCM^\tinyLATMIN \le h_\tinyPCMKCero(t\!+\!\Delta t) \le h_\tinyPCM^\tinyLATMAX$. That means that there is no change in the enthalpic state of the layers interior to $k_0$, as described in Equation \eqref{eq:h_pcm_k_forward}.

        \begin{equation}
            h_\tinyPCMK(t\!+\!\Delta t) \,=\, h_\tinyPCMK(t) \qquad \forall\, k<k_0 \label{eq:h_pcm_k_forward}
        \end{equation}

        	However, layers exterior to $k_0$ are in sensible zone and their enthalpic state needs to be updated. First, the temperature is estimated as shown in Equation \eqref{eq:T_pcm_k_forward}, applying steady-state conduction, where $R_\tinyPCMK^\tinyCOND(t)$ is computed in a similar way to $R_\tinyPCMKCero^\tinyCOND(t)$ in step 2.

        \begin{equation}
            T_\tinyPCMK(t\!+\!\Delta t) \,=\, T_\tinyPCMKMenos(t\!+\!\Delta t) + \dot Q_\tinyPCMKCero(t)\, R_\tinyPCMK^\tinyCOND(t)\qquad \forall\,k>k_0
            \label{eq:T_pcm_k_forward}
        \end{equation}

        	Then, depending on the particular sensible zone, the enthalpic state of the layers exterior to $k_0$ is updated, as shown in Equation Set \eqref{eq:h_pcm_k_forward_exterior}.

        \begin{equation}
            \begin{aligned}
            h_\tinyPCMK(t\!+\!\Delta t) &= h_\tinyPCM^\tinyLATMIN - c_{\!p_\tinyPCMSOL} \! \left(T_\tinyPCM^\tinyLAT - T_\tinyPCMK(t\!+\!\Delta t)\right) \;\; &\scriptstyle{\text{if $k$ is subcooled layer}} \\
            h_\tinyPCMK(t\!+\!\Delta t) &= h_\tinyPCM^\tinyLATMAX + c_{\!p_\tinyPCMLIQ} \! \left(T_\tinyPCMK(t\!+\!\Delta t) - T_\tinyPCM^\tinyLAT\right) \;\; &\scriptstyle{\text{if $k$ is superheated layer}}
            \end{aligned}
            \label{eq:h_pcm_k_forward_exterior}
        \end{equation}

 			The cooling power the whole capsule exchanges with the intermediate fluid, $\dot Q_\tinyPCM(t)$, matches the cooling power exchanged between the intermediate fluid and $k_0$ layer, $\dot Q_\tinyPCMKCero(t)$. Eventually, the temperature of the intermediate fluid is updated using the discrete approximation given in Equation \eqref{eqPrincipalDobleDiscreta}.
 			
        \item
         	Layer $k_0$ quits latent zone. That means that latent energy of layer $k_0$ depleted some time before the period expired, $\Delta t_\tinyKCero \le \Delta t$, when the layer entered sensible zone. The precise instant is estimated, considering a constant rate, $\dot Q_\tinyPCMKCero$, and the remaining latent energy of layer $k_0$, as shown in Equation \eqref{eq:Deltat_k0_Charging} for a charging process or Equation \eqref{eq:Deltat_k0_Discharging} for a discharging process.
         	
        \begin{equation}
           \Delta t_{k_0} = \frac{\rho_\tinyPCMKCero(t) V_\tinyPCMKCero(t) (h_\tinyPCM^\tinyLATMIN - h_\tinyPCMKCero(t))} {\dot Q_\tinyPCMKCero(t)}
           \label{eq:Deltat_k0_Charging}
        \end{equation}
        
        \begin{equation}
           \Delta t_{k_0} = \frac{\rho_\tinyPCMKCero(t) V_\tinyPCMKCero(T) (h_\tinyPCM^\tinyLATMAX - h_\tinyPCMKCero(t))} {\dot Q_\tinyPCMKCero(t)}
           \label{eq:Deltat_k0_Discharging}
        \end{equation}
        
			Then, a shorter time interval, $\Delta t_\tinyKCeroMenos$, is transiently fixed, as indicated in Equation \eqref{eq:Deltat_k0}.

        \begin{equation}
            \Delta t_\tinyKCeroMenos \,=\, \Delta t - \Delta t_\tinyKCero
            \label{eq:Deltat_k0}
        \end{equation}
        
        	This shorter time interval is used for the residual estimation, for which the next inner layer, $k_0\!-\!1$, is established as the new first layer in latent zone, and the sequence restarts from step 2, but using the corresponding residual interval, $\Delta t_\tinyKCeroMenos$, instead of $\Delta t$.

        	This way, the cooling power transferred between each PCM capsule and the intermediate fluid, $\dot Q_\tinyPCM(t)$ has eventually several successive contributions:
        	
        \[ \dot Q_\tinyPCM(t) \,=\, \left\{ \dot Q_\tinyPCMKCero(t),\, \dot Q_\tinyPCMKCeroMenos(t),\, \ldots \dot Q_\tinyPCMKCeroMenosJ(t)\right\}\; , \]

        	being $k\!=\!k_0\!-\!j$ the most inner layer found in latent state during period $\Delta t$. Clearly, the addition of the partial intervals amounts for the whole period, as shown in Equation \eqref{eq:Deltat_sumatorio}.

        \begin{equation}
            \sum_{i=0}^{i=j} \Delta t_\tinyKCeroMenosI \,=\, \Delta t
            \label{eq:Deltat_sumatorio}
        \end{equation}

        	For each one of these partial intervals, the temperature of the intermediate fluid is updated using Equation \eqref{eqPrincipalDobleDiscreta}:

        \begin{equation*}
            \begin{split}
            T_\tinyINT(t+\Delta t_\tinyKCero) \,=\,
            T_\tinyINT(t) - \frac{\Delta t_\tinyKCero}{m_\tinyINT\,c_{\!p_\tinyINT}} \,
            \Big [
             n_\tinyREF \, \dot Q_\tinyREF(t) + n_\tinySEC \, \dot Q_\tinySEC(t) + n_\tinyPCM \, \dot Q_\tinyPCMKCero(t)
            \Big ]
            \end{split}
        \end{equation*}

        \begin{equation*}
            \hspace{-3em}
            T_\tinyINT(t+\Delta t_\tinyKCeroMenos) \,=\, T_\tinyINT(t+\Delta
            t_\tinyKCero) - \frac{\Delta t_\tinyKCeroMenos}{m_\tinyINT\,c_{\!p_\tinyINT}} \,
            \Big [
            n_\tinyREF \, \dot Q_\tinyREF(t) + n_\tinySEC \, \dot Q_\tinySEC(t) + n_\tinyPCM \, \dot
            Q_\tinyPCMKCeroMenos(t)
            \Big ]
        \end{equation*}

        \begin{equation*}
            \cdots
        \end{equation*}

    \end{itemize}

\end{enumerate}


\section{Accuracy assessment of the time-efficient discrete model}\label{secAccuracyAssessment}

The original discrete model and the proposed "accelerated" version are compared in this section. Simulations covering full charging/discharging processes, as well as sequences of partial charging and discharging operations are presented, allowing to visualise the interesting effects taking place. The efficient discrete model is run using different time periods: $\Delta t =\{60\, \text{s},\,180\, \text{s},\,300\, \text{s}\}$, while the number of layers has been fixed to $n_\tinyLAY=10$.

\subsection{Evaluation under full charging operation}\label{secFullCharginOperation}

\begin{figure}[h]
    \centerline{\includegraphics[width=12.0cm,angle=0]{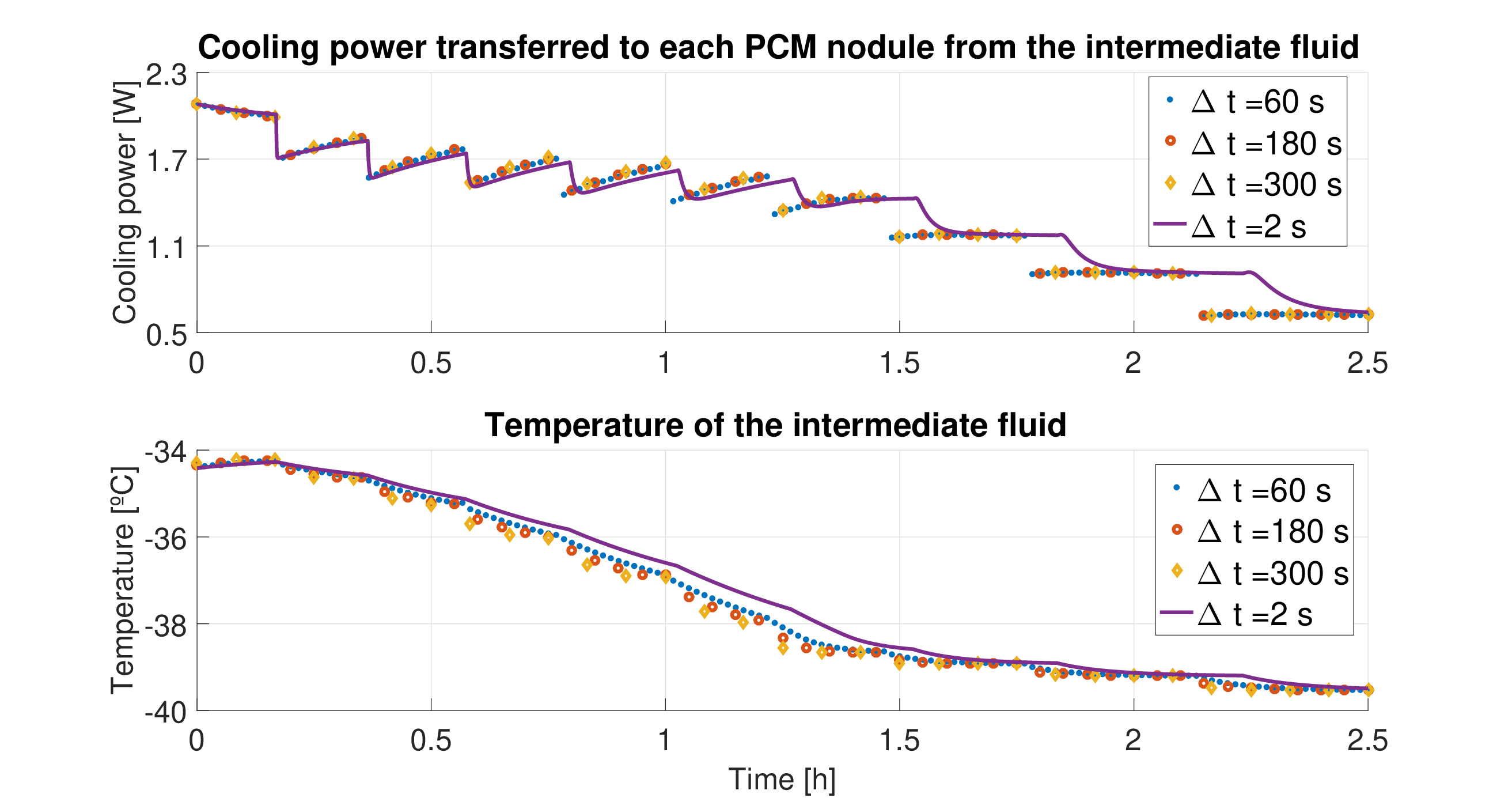}}
    \caption{Comparison of the transferred cooling power between each PCM nodule and the intermediate fluid (upper plot) and its temperature (lower plot), in a typical charging operation.}
    \label{figComparisonDiscrDiscr_PowerIntermFluidOneCapsulePCM_charge}
\end{figure}

First, the different discrete models are compared during a full charging simulation. Figure \ref{figComparisonDiscrDiscr_PowerIntermFluidOneCapsulePCM_charge} represents the comparative evolution of the cooling power transferred to each PCM nodule from the intermediate fluid, during a typical charging process, and the temperature of the intermediate fluid. The original discrete model is represented by continuous line and denoted by $\Delta t = 2\, \text{s}$ in the plot legend, whereas the dotted line refers to the accelerated discrete model with $\Delta t = 60\, \text{s}$, the circles correspond to $\Delta t = 180\, \text{s}$ and, finally, the diamonds are for $\Delta t=300\, \text{s}$.

\begin{figure}[h]
	\centerline{\includegraphics[width=12.0cm,angle=0]{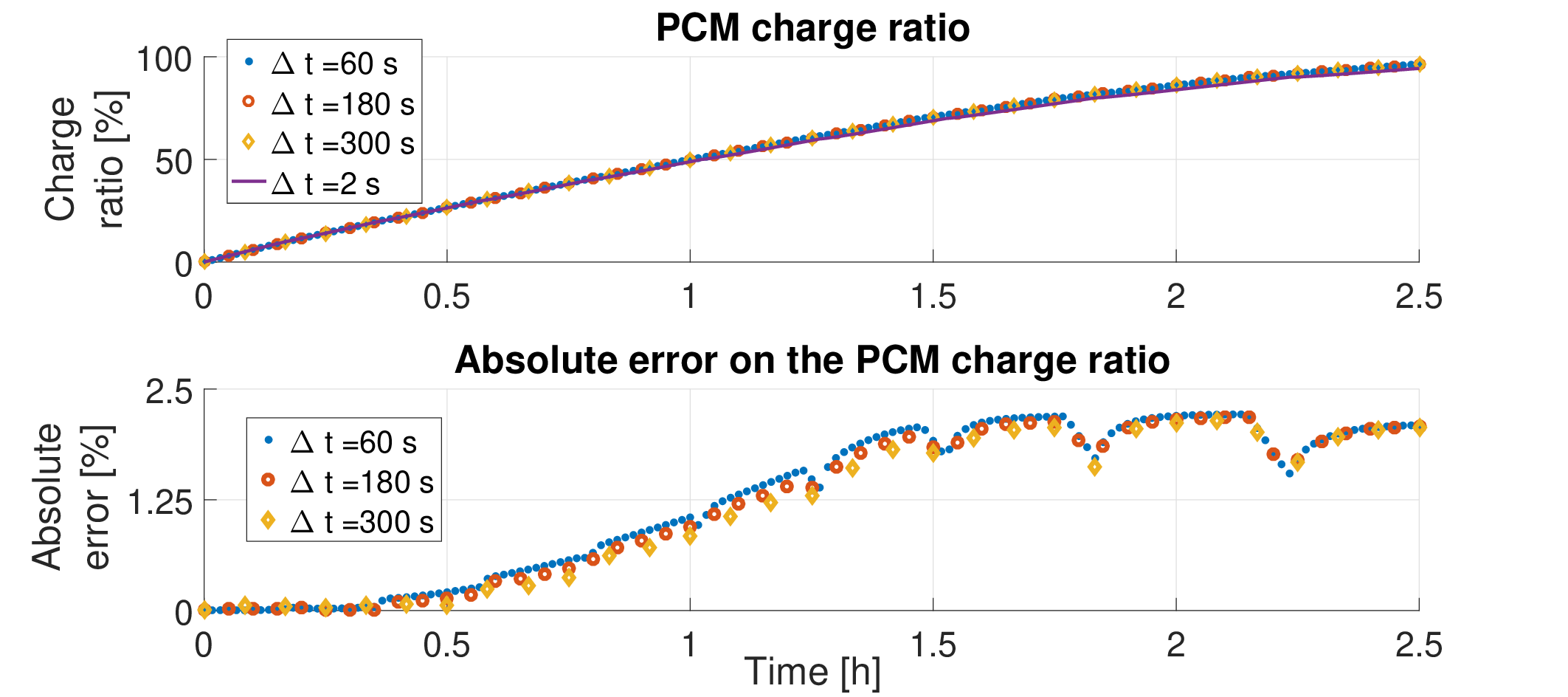}}
	\caption{Comparison of the PCM charge ratio during the simulated typical charging operation.}
	\label{figComparisonDiscrDiscrDiscr_ChargeRatioError_charge}
\end{figure}

Figure \ref{figComparisonDiscrDiscrDiscr_ChargeRatioError_charge} gives the comparison of the PCM charge ratio, for the time-efficient discrete models, with respect to the original discrete one. The charge ratio is deemed as the best quantitative way to assess the accuracy of the efficient model, given the proposed use of this model within energy management strategies. According to this, the corresponding absolute error in the estimation of the charge ratio is represented in the lower plot of Figure \ref{figComparisonDiscrDiscrDiscr_ChargeRatioError_charge}.

\begin{figure}[H]
    \centerline{\includegraphics[width=12.0cm,angle=0]{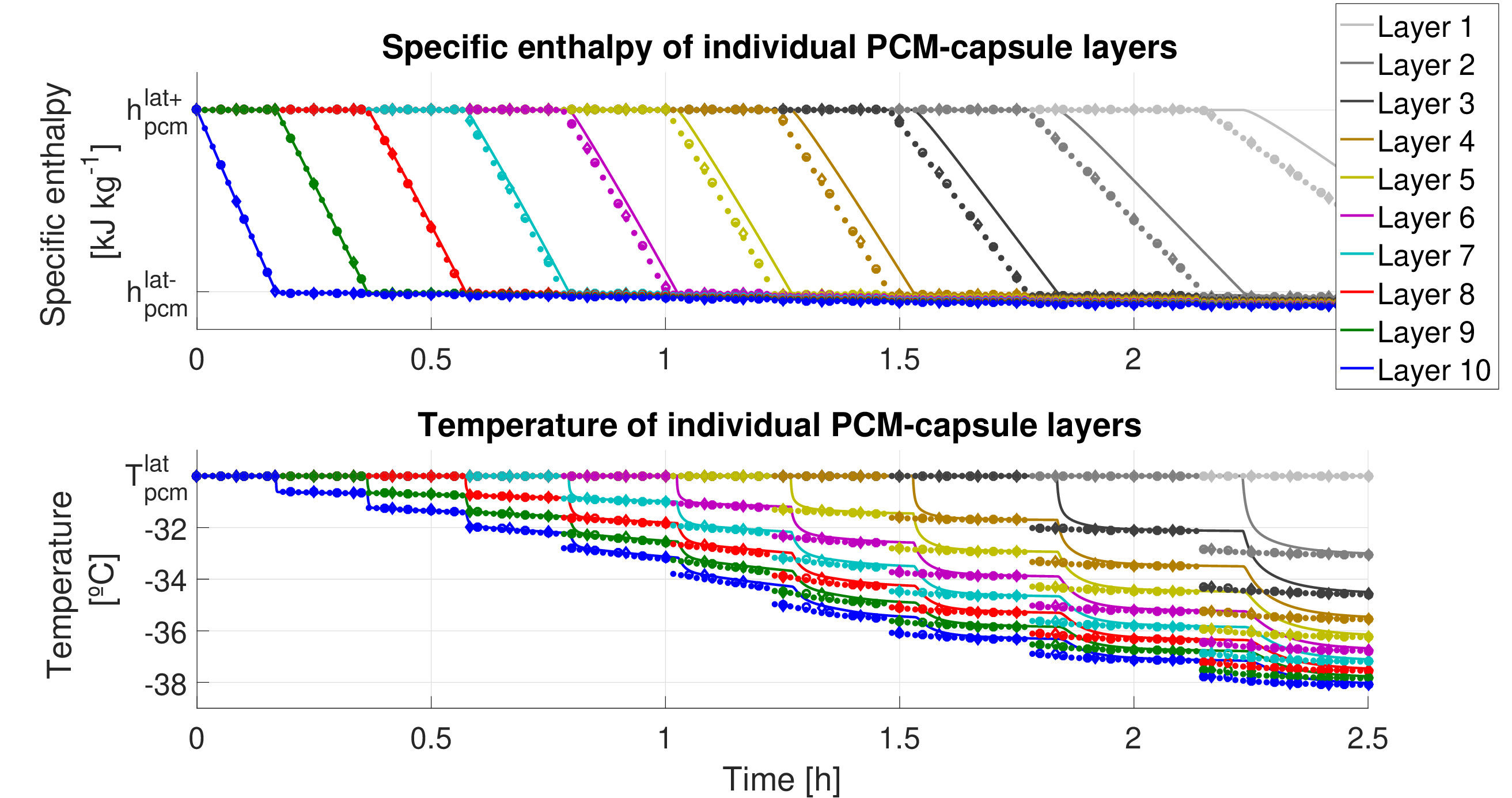}}
    \caption{Evolution of specific enthalpy (upper plot) and temperature (lower plot) of the individual PCM-nodule layers during the simulated charging operation.}
    \label{figComparisonDiscrDiscr_EntalpiaTemperaturaCapas_charge}
\end{figure}

It is interesting to analyse the time evolution of the specific enthalpy and temperature of the individual layers during the simulated charging operation. It can be seen in Figure \ref{figComparisonDiscrDiscr_EntalpiaTemperaturaCapas_charge}, where colours have been allocated to highlight the different layers. It can be noticed that even the most scarcely sampled case ($\Delta t = 300\, \text{s}$) is able to suitably track the dynamic evolution of each individual layer.

Recalling that one of the main assumptions made to develop the time-efficient model is that the dynamics of the layers in sensible zone are considered as negligible,
calculating their temperature according to a steady-state conduction model. This is not a weak assumption, given that the temperature dynamics of the layers in sensible zone are far away from being instantaneous, as appreciated in Figure \ref{figComparisonDiscrDiscr_EntalpiaTemperaturaCapas_charge}, regarding the continuous lines describing the performance of the original discrete model. However, in view of Figure \ref{figComparisonDiscrDiscr_EntalpiaTemperaturaCapas_charge}, it is clear that the dynamics related to the layers in latent zone are slower than those of the layers in sensible zone, due to the continuous change in their temperature. The most realistic model is undoubtedly the discrete one, which describes the microscopic truth in detail, but given the aimed use of this efficient model, calculation speed turns out to be a priority and lower computation time can only be achieved by assuming a certain degree of inaccuracy. In fact, when designing the energy management strategies, we are mainly interested on the macroscopic charge ratio of the whole TES tank, not so much on the microscopic details. As shown in Figure \ref{figComparisonDiscrDiscrDiscr_ChargeRatioError_charge}, the absolute errors regarding the charge ratio remain under 2.5\% in this full charging simulation, even when considering the longest $\Delta t$. Given the achieved reduction on the calculation time, which will be detailed in Section \ref{secComputationalCost}, and particularly the aimed use of this time-efficient model, this level of inaccuracy represents a very reasonable trade-off. 

\begin{figure}[h]
    \centerline{\includegraphics[width=13cm,trim = 0 0 740 0,clip]
    {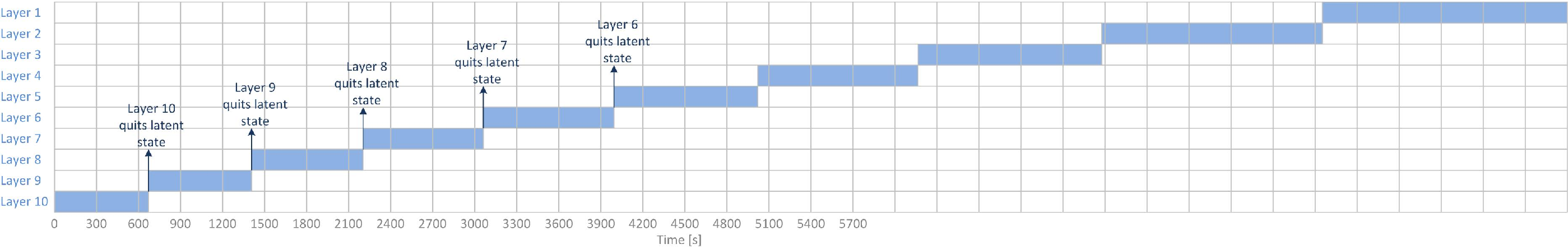}}
    \caption{Timing diagram showing latent phase progression during a full charge, ten-layer example, with $\Delta t = 300$~s.}
    \label{figDeltaTCarga}
\end{figure}

Finally, Figure \ref{figDeltaTCarga} is given as a further illustration of how the proposed algorithm works timewise. It corresponds to the full charging experiment just described, in particular, using $\Delta t = 300$~s. This diagram displays the intervals when each one of the set of ten layers evolves through latent phase, emphasizing the exact instant at which a layer quits latent phase and the next one becomes the outermost layer evolving through latent phase. We can appreciate that the base time period $\Delta t$ is split when the algorithm predicts latent-phase evolution shifts from one layer to the next. Certainly, this is one of the simplest cases we could provide. In fact, in this case, we find no more than one single "jump" in between two consecutive major time steps. If 50 layers had been chosen instead, for instance, we could certainly see several "jumps" inside the same base time period. On the other hand, had the experiment not been started from a fully discharged state, but after an arbitrary sequence of partial charging and discharging operations, we could see more transitions inside the same base time period, due to the fact that the enthalpy of some layers could have already been half way through its latent interval.

\subsection{Evaluation under full discharging operation}

\begin{figure}[H]
    \centerline{\includegraphics[width=12.0cm,angle=0]{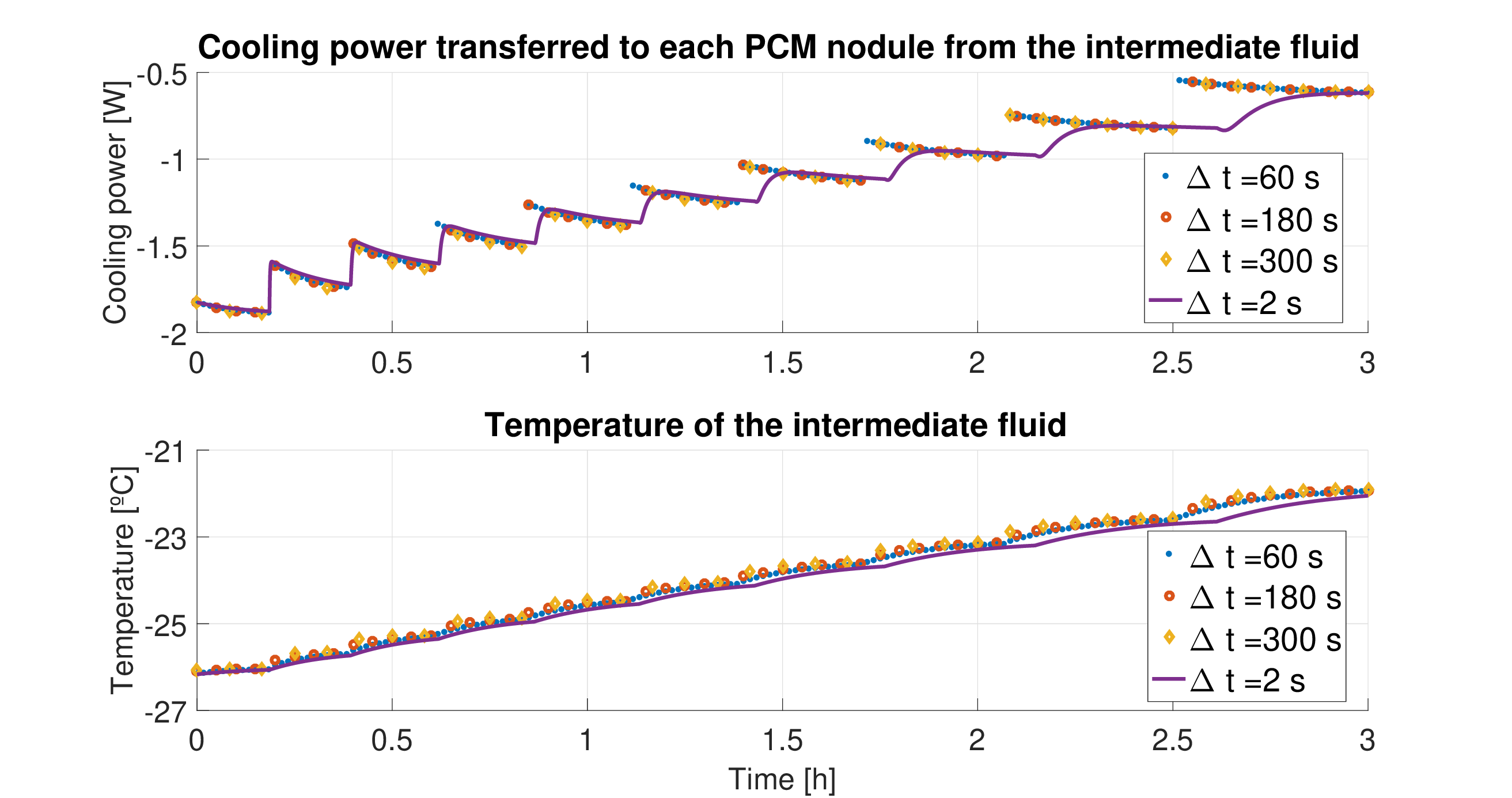}}
    \caption{Comparison of the transferred cooling power between each PCM nodule and the intermediate fluid (upper plot) and its temperature (lower plot), in a typical discharging operation.}
    \label{figComparisonDiscrDiscr_PowerIntermFluidOneCapsulePCM_discharge}
\end{figure}

\begin{figure}[h]
	\centerline{\includegraphics[width=12.0cm,angle=0]{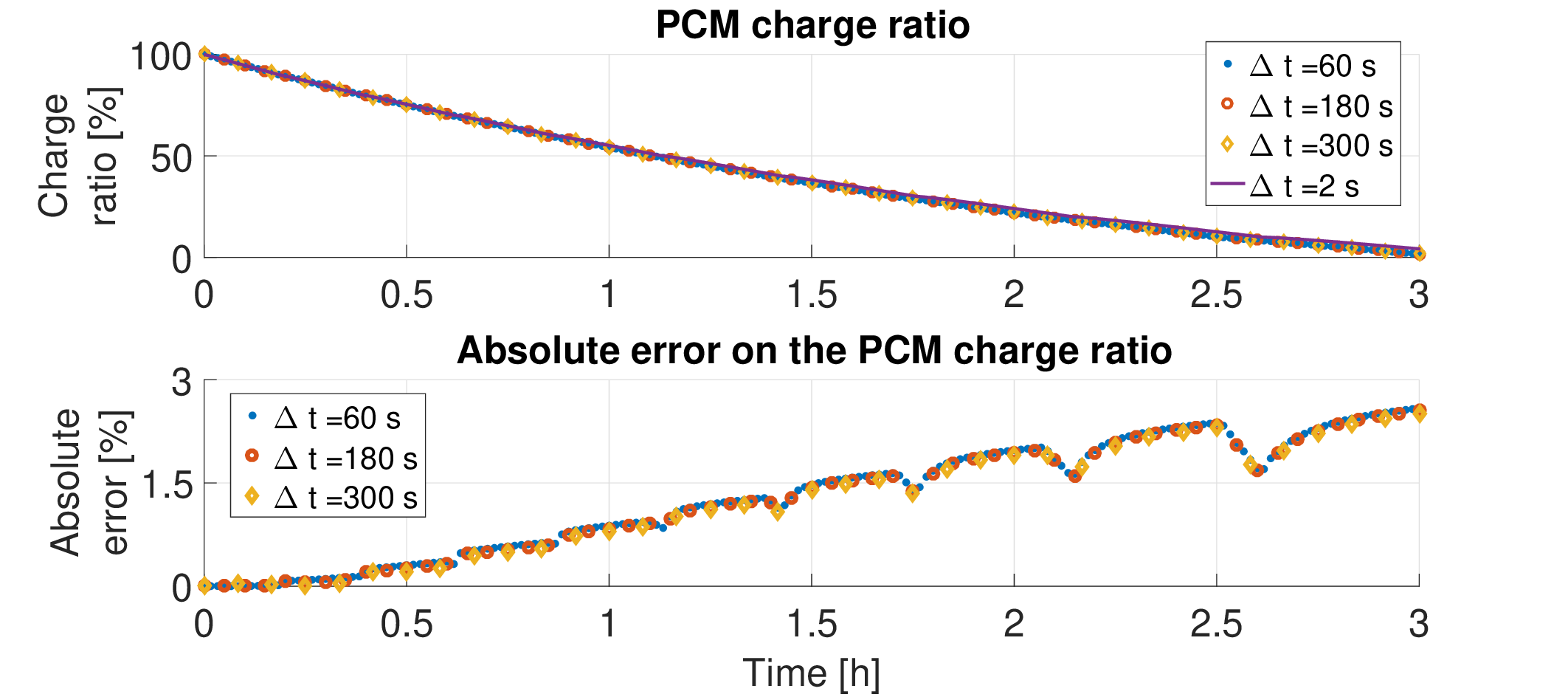}}
	\caption{Comparison of the PCM charge ratio during the simulated typical discharging operation.}
	\label{figComparisonDiscrDiscrDiscr_ChargeRatioError_discharge}
\end{figure}

Secondly, the different discrete models are compared during a full discharging simulation. Figure \ref{figComparisonDiscrDiscr_PowerIntermFluidOneCapsulePCM_discharge} represents the comparative evolution of the cooling power transferred from each PCM nodule to the intermediate fluid, during a typical discharging process, and the temperature of the intermediate fluid. Furthermore, Figure \ref{figComparisonDiscrDiscrDiscr_ChargeRatioError_discharge} gives the comparison of the PCM charge ratio, as well as the corresponding absolute error in the estimation of this ratio.

It can be seen in Figure \ref{figComparisonDiscrDiscrDiscr_ChargeRatioError_discharge} that the discrepancy between the simplified discrete model and the original remains under 3\% for all sampling periods along the complete discharge, similar to the errors shown in Figure \ref{figComparisonDiscrDiscrDiscr_ChargeRatioError_charge}.

\begin{figure}[H]
    \centerline{\includegraphics[width=12.0cm,angle=0]{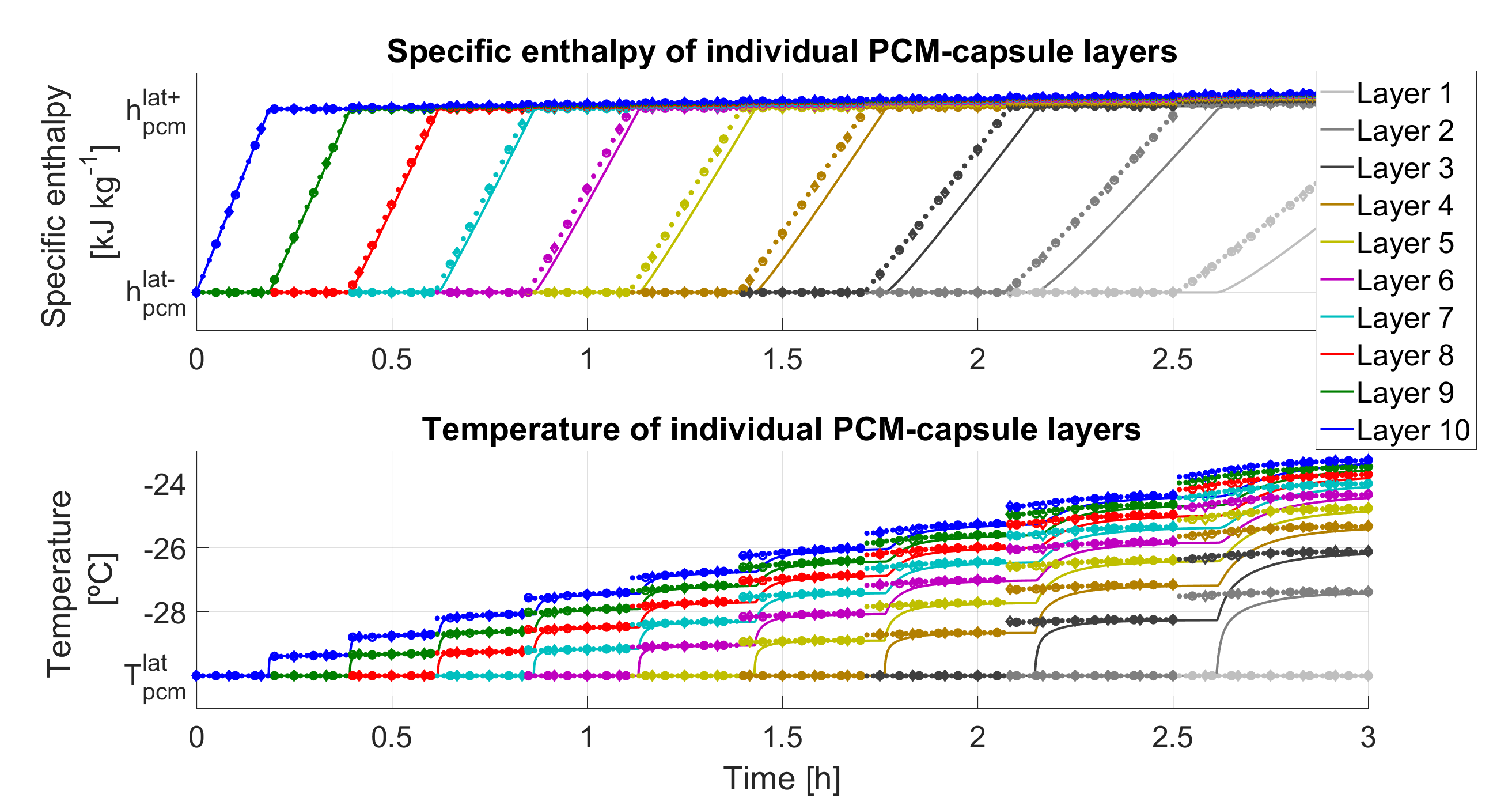}}
    \caption{Evolution of specific enthalpy (upper plot) and temperature (lower plot) of the individual PCM-nodule layers during the simulated discharging operation.}
    \label{figComparisonDiscrDiscr_EntalpiaTemperaturaCapas_discharge}
\end{figure}

Additionally, the time evolution of the specific enthalpy and temperature of the individual layers during the simulated discharging operation is represented in Figure
\ref{figComparisonDiscrDiscr_EntalpiaTemperaturaCapas_discharge}.

\subsection{Evaluation under series of partial charging/discharging operations}

\begin{figure}[H]
    \centerline{\includegraphics[width=12.0cm,angle=0]{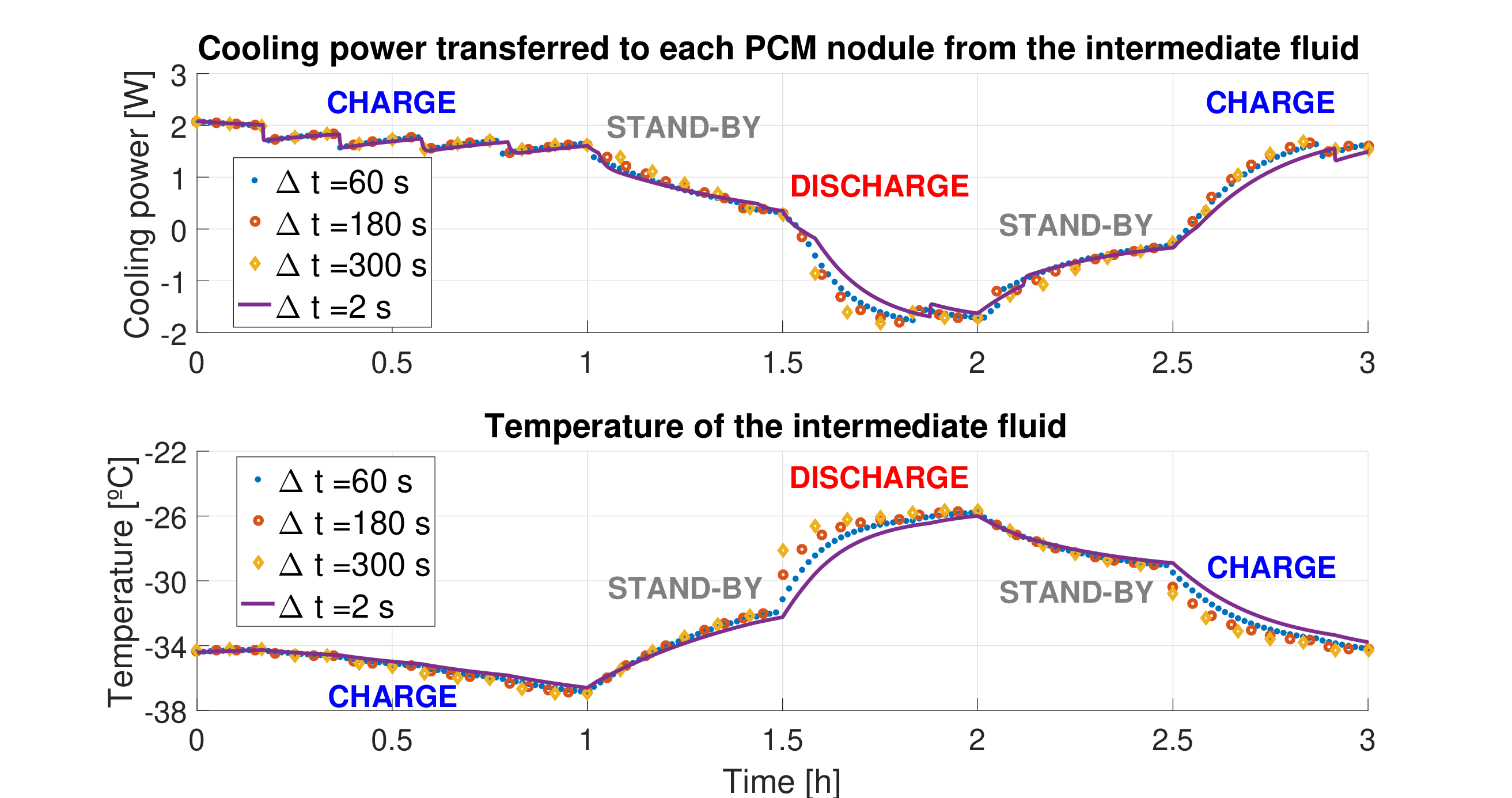}}
    \caption{Comparison of the transferred cooling power between each PCM nodule and the intermediate fluid (upper plot), and its temperature (lower plot), in a series of charging/discharging operations.}
    \label{figComparisonDiscrDiscr_PowerEnergyOneCapsulePCM_MultiplesCargasDescargas}
\end{figure}

\begin{figure}[H]
	\centerline{\includegraphics[width=12.0cm,angle=0]{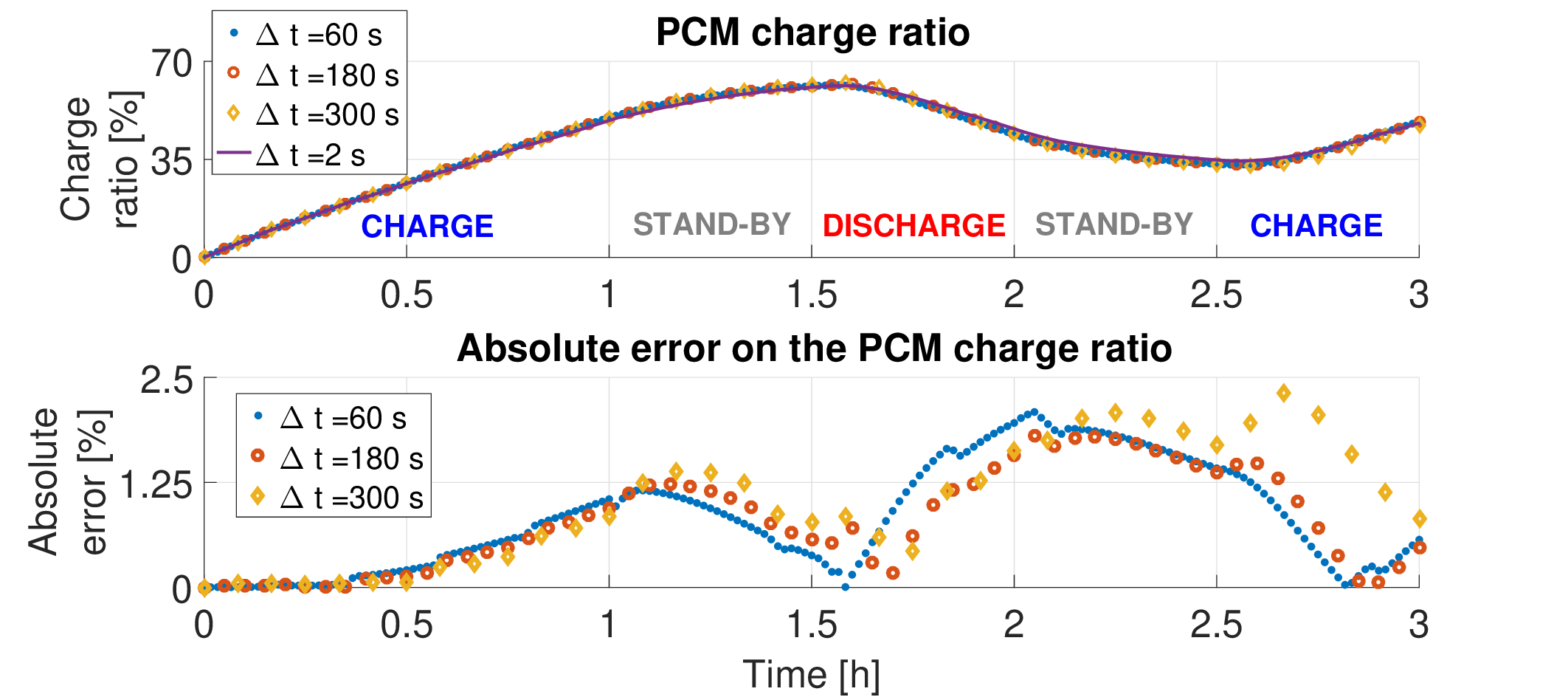}}
	\caption{Comparison of the PCM charge ratio during a series of charging/discharging operations.}
	\label{figComparisonDiscrDiscr_ChargeRatioError_MultiplesCargasDescargas}
\end{figure}

\begin{figure}[H]
    \centerline{\includegraphics[width=12.0cm,angle=0]{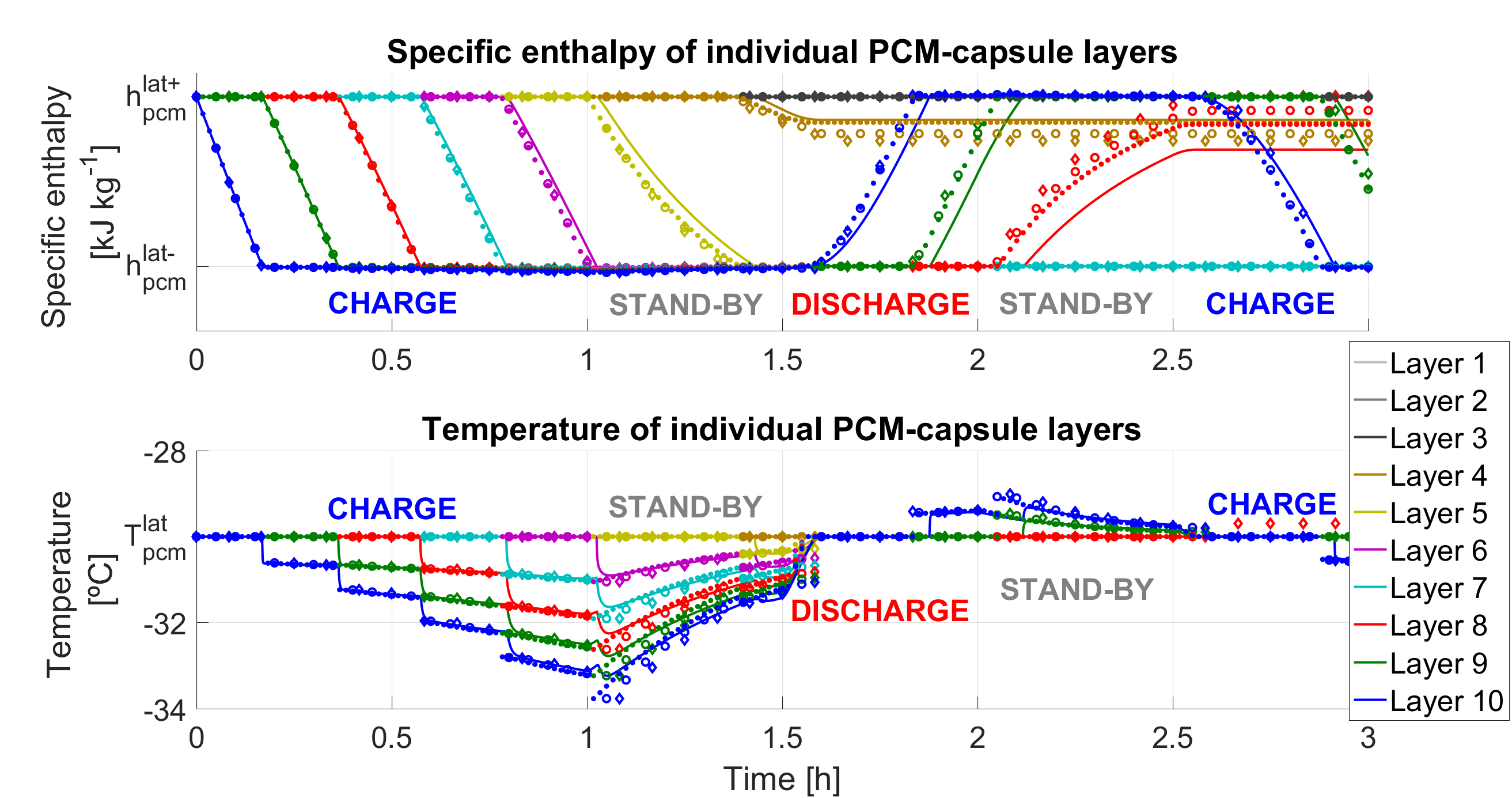}}
    \caption{Evolution of specific enthalpy (upper plot) and temperature (lower plot) of the individual PCM-capsule layers during a series of charging/discharging operations.}
    \label{figComparisonDiscrDiscr_EntalpiaTemperaturaCapas_MultiplesCargasDescargas}
\end{figure}

Next, a more complex scenario is assumed. A series of partial charging/discharging operations with a short stand-by period is simulated. The same set of plots as for  the full charging/discharging operations is provided in Figures \ref{figComparisonDiscrDiscr_PowerEnergyOneCapsulePCM_MultiplesCargasDescargas} -- \ref{figComparisonDiscrDiscr_EntalpiaTemperaturaCapas_MultiplesCargasDescargas}. The particular sequence is as follows: 1 hour charge, 0.5 hours stand-by, 0.5 hours discharge, 0.5 hours stand-by and 0.5 hours charge; the PCM nodules are considered to be initially completely discharged.

Figure \ref{figComparisonDiscrDiscr_EntalpiaTemperaturaCapas_MultiplesCargasDescargas} is very interesting, as it can be observed that, during the first one-hour charging period, there is time for the five most external layers to reach minimum enthalpy levels. During the next half-hour stand-by period, however, due to thermal inertia, layer 5 starts to lower its enthalpy level, leaving latent zone and allowing also layer 4 to reduce its specific enthalpy. After starting the next half-hour discharging operation, there is only time for the two most external layers to gain enthalpy. During the next half-hour stand-by period, layer 9 leaves latent zone and layer 8 has time to gain enthalpy by thermal inertia, whereas during the final charging period starts, only layer 10 leaves latent zone.

\vspace{1em} 

All these simulations reveal that the absolute errors between the simplified and the original discrete model remain below 3\% concerning the PCM charge ratio, which turns out to be the most interesting variable to estimate the cold energy stored within the TES tank. This information is very valuable for high-level energy management strategies to schedule the charge/discharge of the TES tank, in order to optimize some economic and energy efficiency criteria while satisfying a given cooling demand. Table \ref{tabErrors} gives further details about the maximum and average errors on the charge ratio for the considered values of $\Delta t$.

\begin{table}[h]
	\scalebox{0.75}[0.75]{ \tabulinesep=0.5mm
		\begin{tabu} { C{2.5cm} C{2cm} C{2cm} C{2cm} C{2cm} C{2cm} C{2cm} }
			\toprule
			  & \multicolumn{2}{c}{$\Delta t\,$= 60 s} & \multicolumn{2}{c}{$\Delta t\,$= 180 s} & \multicolumn{2}{c}{$\Delta t\,$= 300 s} \\ 
			 \cmidrule{2-7}
			  & Maximum error [\%] & Average error [\%] & Maximum error [\%] & Average error [\%] & Maximum error [\%] & Average error [\%] \\ 
			 \midrule
			 \bf{Charge} & 2.21 & 1.27 & 2.18 & 1.22 & 2.14 & 1.16 \\ 
			 \midrule
			 \bf{Discharge} & 2.59 & 1.31 & 2.55 & 1.28 & 2.51 & 1.24 \\ 
			 \midrule
			 \bf{Partial operations} & 2.09 & 0.83 & 1.80 & 0.79 & 2.31 & 1.01 \\ 
			 \bottomrule
		\end{tabu}}
	\caption{Maximum and average errors on the PCM charge ratio for the considered $\Delta t$ and simulation cases.}
	\label{tabErrors}
\end{table}


\section{Comparative computational cost evaluation} \label{secComputationalCost}

The computational cost of the different discrete implementations of the model has been evaluated and some conclusions are presented in this section. The hardware/software configuration under which the simulations have been performed is the following: Windows 8.1 64 bits, CPU Intel Core i7-3632QM 2.20 GHz, featuring 8 GB RAM. MATLAB R2013a 64 bits has been used as simulation software. Note that the point of this section is to emphasize the reduction of the simulation time in relative terms.

\begin{figure}[H]
    \centering
    \subfigure[Full charging simulation]{
        \includegraphics[width=7.0cm,angle=0]{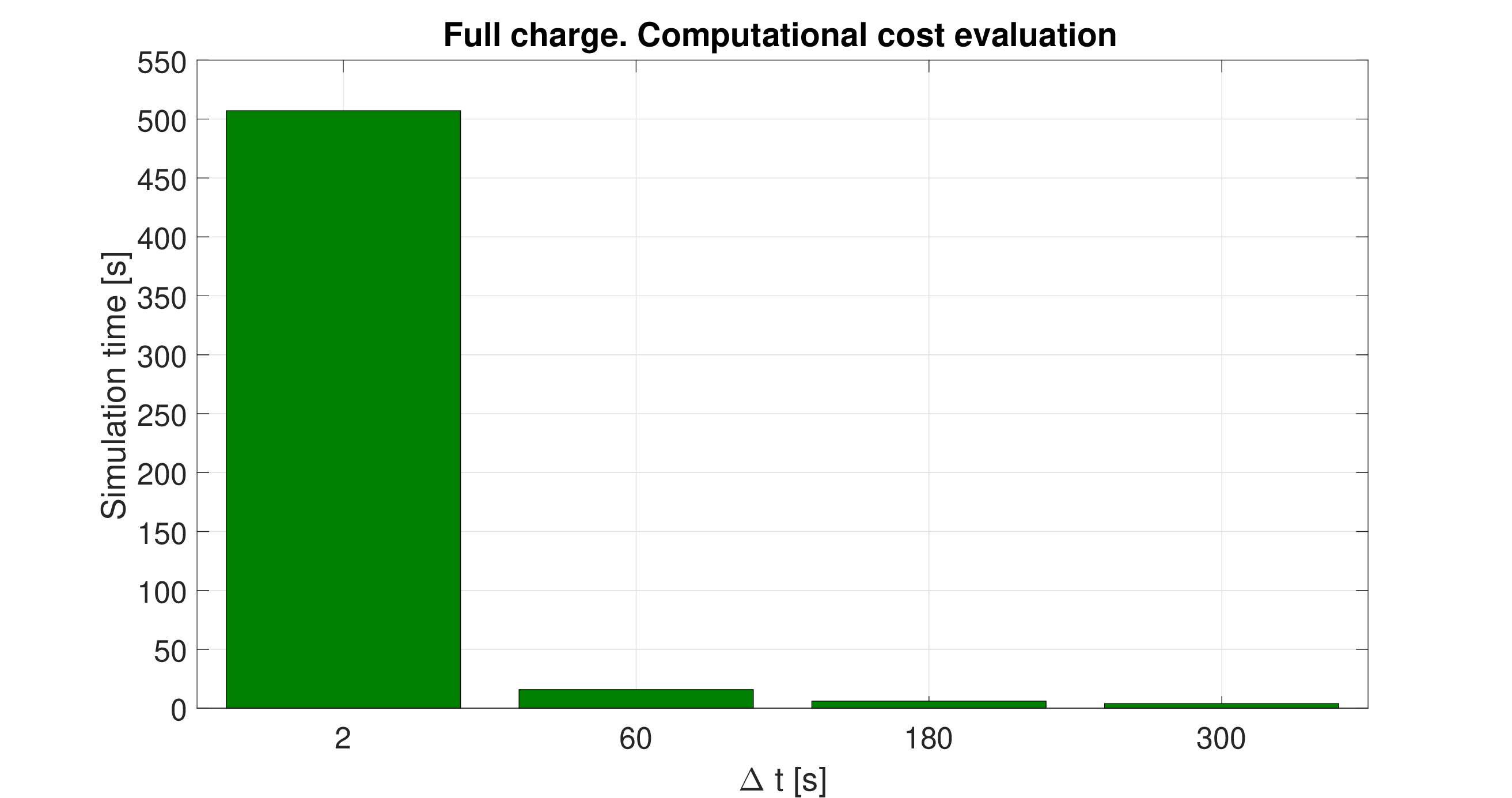}
        \label{fig_computacional_cost_charge}
    }\subfigure[Full discharging simulation]{
        \includegraphics[width=7.0cm,angle=0]{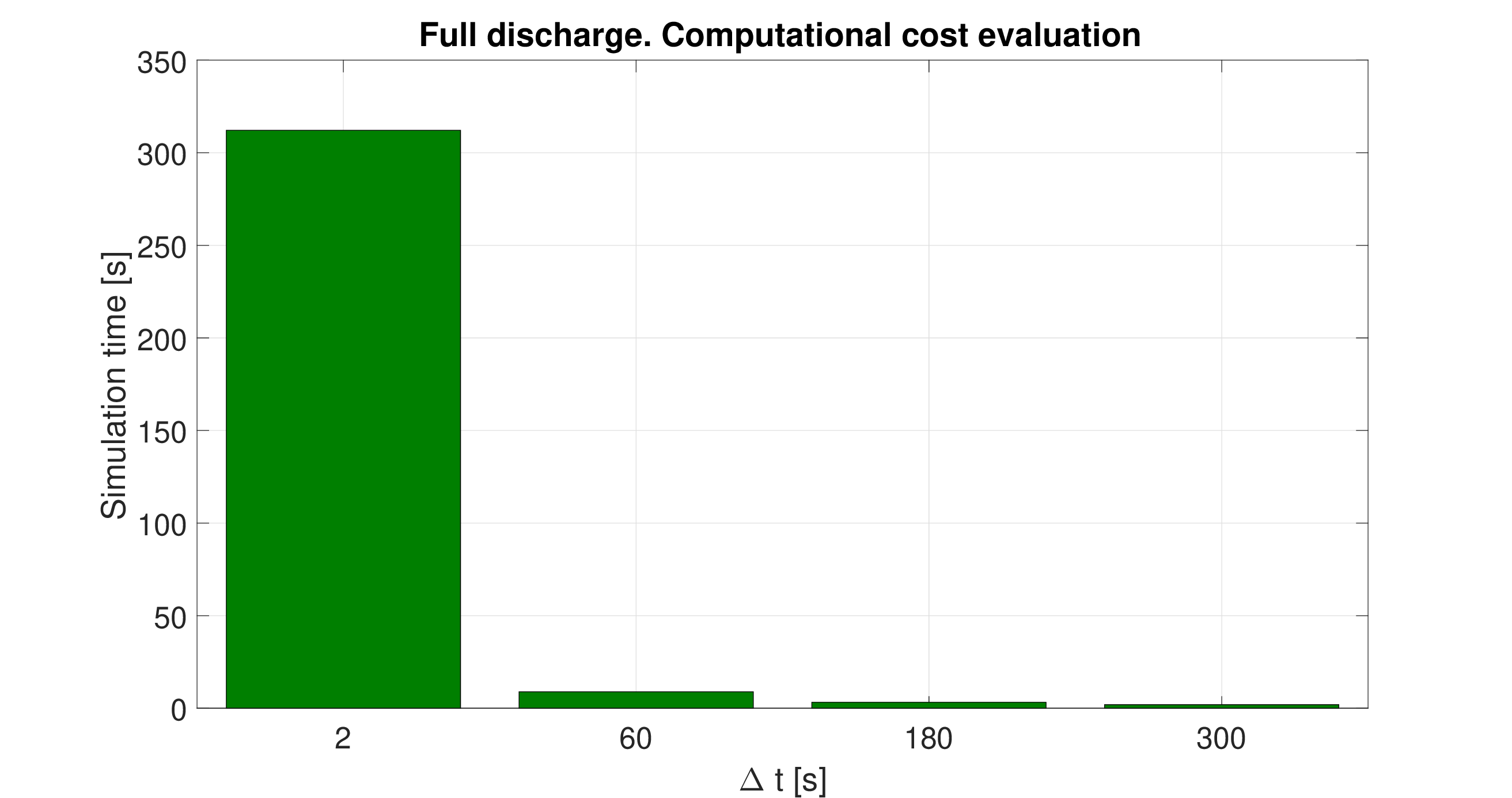}
        \label{fig_computacional_cost_discharge}
    }
    \subfigure[Partial charging/discharging sequence]{
        \includegraphics[width=7.0cm,angle=0]{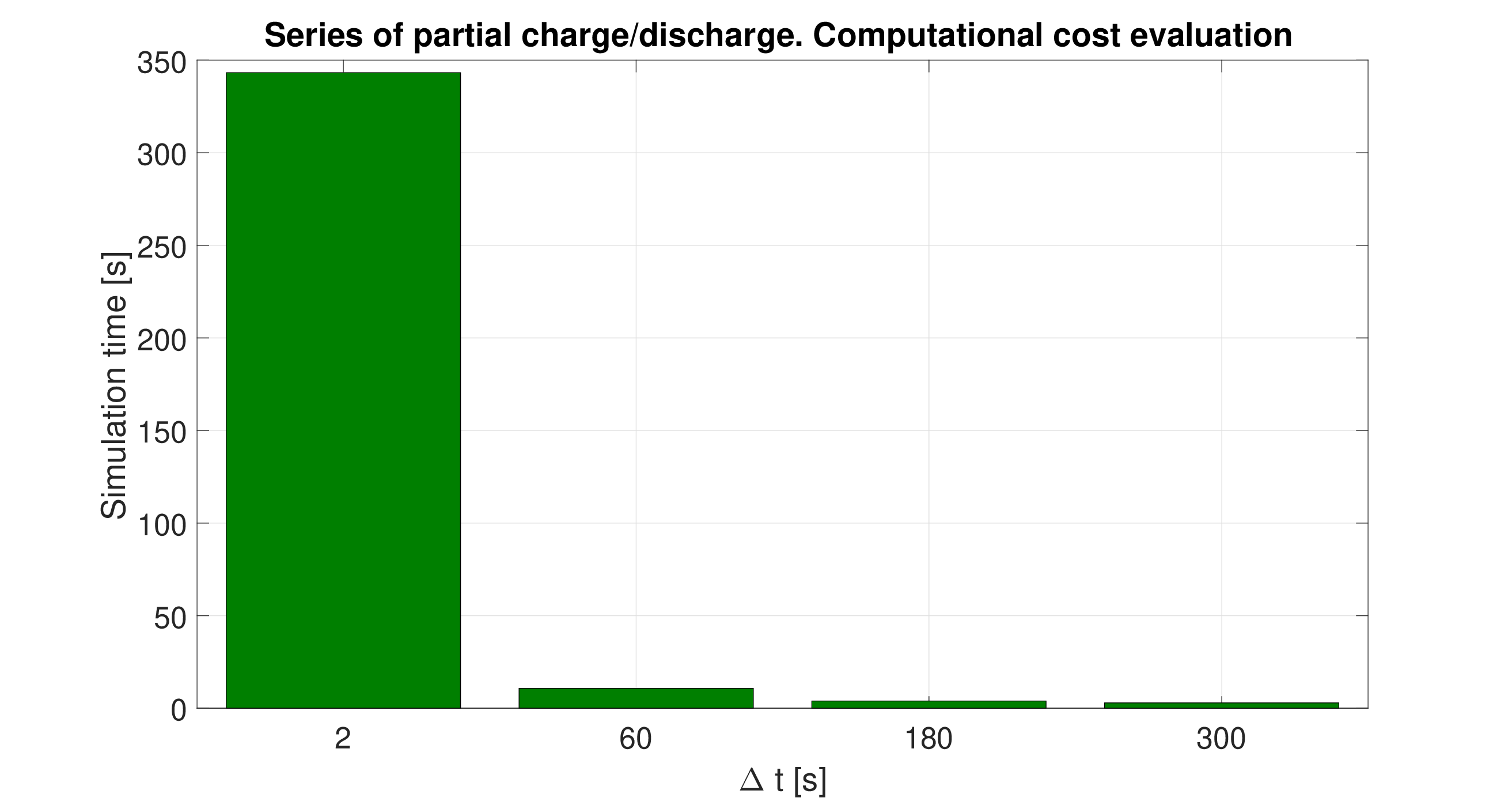}
        \label{fig_computacional_cost_multiple}
    }
    \caption{Comparative computational cost evaluation of the discrete approximations of the PCM-based cold-energy storage model.}
    \label{figComputationalCostFullChargeSeriesPartial}
\end{figure}

First, the simulation of the full charging operation, amounting for a total of 2.5 real-time hours, is analysed. The comparative computational cost evaluation of the different discrete approximations is given in the left part of Table \ref{tabComputationalCostFullChargeSeriesPartial}. Secondly, the computational cost of the 3-hour full discharging operation is detailed in the central columns of the table. Finally, the computational workload of the 3-hour simulation of the series of partial charging/discharging operations is evaluated in the right part of the table. Figure \ref{figComputationalCostFullChargeSeriesPartial} reproduces the same information in a graphical way.

\begin{table}[h]
    \scalebox{0.75}[0.75]{ \tabulinesep=0.5mm
        \begin{tabu} { L{1.3cm} L{3.5cm} }
            \multicolumn{2}{c}{\bf Full charging operation} \\
            $\Delta t\,$[s] & Simulation time $[\text{s}]$ \\ \hline
            $2$ & $507.1368$ \\ \hline
            $60$ & $15.8681$ \\ \hline
            $180$ & $6.1116$ \\ \hline
            $300$ & $3.9998$ \\
        \end{tabu}
        \begin{tabular} {L{1.3cm} L{3.5cm} }
            \multicolumn{2}{c}{\bf Full discharging operation}\\
            $\Delta t\,$[s] & Simulation time $[\text{s}]$ \\ \hline
            $2$ & $312.1315$ \\ \hline
            $60$ & $8.9100$ \\ \hline
            $180$ & $3.2213$ \\ \hline
            $300$ & $1.9465$ \\
        \end{tabular}
        \begin{tabu} {L{1.3cm} L{3.5cm} }
            \multicolumn{2}{c}{\bf Partial charging/discharging}\\
            $\Delta t\,$[s] & Simulation time $[\text{s}]$ \\ \hline
            $2$ & $343.2252$ \\ \hline
            $60$ & $10.7955$ \\ \hline
            $180$ & $3.9215$ \\ \hline
            $300$ & $2.9326$ \\
    \end{tabu}}
    \caption{Simulation times required by the different discrete approximations of the PCM-based cold-energy storage model.}
    \label{tabComputationalCostFullChargeSeriesPartial}
\end{table}

Table \ref{tabComputationalCostFullChargeSeriesPartial} reveals a drastic reduction on the required simulation time. In particular, simulation runs about 32 times faster using $\Delta t = 60\,\text{s}$, around 83 times faster when using $\Delta t = 180\,\text{s}$, and over 127 times faster, in the case $\Delta t = 300\,\text{s}$. These seem to be very appealing numbers, provided that, as mentioned earlier, the errors on the PCM charge ratio remain under 3\% in every case. These results are consistent with many other simulations performed on the TES system, including non-nominal operating conditions regarding the refrigeration cycle and the secondary fluid, where model accuracy and computational time reduction have been studied and very similar quantitative results have been obtained. In particular, it is worth noting that some tests considering up to 50 layers have been performed. Such a number of layers implies that the fixed integration step required by the original discrete model must be reduced enough to properly describe the faster transient related to the layers in sensible zone, which increases exponentially the simulation time. However, the time-efficient discrete model allows to work with longer time periods such as those considered in Section \ref{secAccuracyAssessment}. The only difference is that the time period is divided in much more intervals, since the latent energy of every layer is smaller. It implies increasing the number of iterations of the sequence described in Section \ref{secEfficientDiscreteModels}, which also increases the overall simulation time, but the reduction with respect to the fixed-step integration of the discrete model is even more drastic than that shown in Figure \ref{figComputationalCostFullChargeSeriesPartial} and Table \ref{tabComputationalCostFullChargeSeriesPartial}.


\section{Conclusions} \label{secConclusions}

In this paper, computationally efficient modelling of TES systems based on PCM has been addressed. The work is focused on a recently proposed layout of PCM-based cold storage systems, conceived to complement an existing refrigeration plant. Taking as keystone a recent work, where both a continuous model and a discrete model have been proposed to represent the thermal behaviour of the PCM nodules in both charging and discharging processes, the computational load of the discrete model has been assessed. The discrete model allows to describe more general conditions than the continuous model, such as series of partial charging/discharging operations. Nonetheless, the computing workload required by the discrete model is too high for the intended purpose. 

A time-efficient version of the discrete model has been proposed to reduce computational load, while keeping the overall accuracy of the original model, by means of loosing the discretisation time, among other factors. The (wider) time interval under consideration is divided into a number of subintervals where cold energy is assumed to be transferred at a constant rate. The length of the subintervals is determined by the remaining energy of the outermost spherical layer in latent zone, while the temperature dynamics of every layer in sensible zone are neglected. The accuracy of the proposed time-efficient model has been assessed by simulating full charging/discharging operations, as well as sequences of partial charging/discharging operations. It is shown that the relative errors between the simplified and the original discrete model remain below 3\%, in terms of the estimated PCM charge ratio, whereas the computational load is reduced by up to 99\%. This drastic reduction on the computational cost enables the integration of the TES model together with the refrigeration cycle model, in order to feasibly address some intended future work, focused on efficient cold-energy management strategies, combining production, storage, and load demand fulfilment. 


\section*{Acknowledgements}

The authors would like to acknowledge MCeI (Grants DPI2015-70973-R and DPI2016-79444-R) for funding
this work.

\section*{References}

\bibliographystyle{IEEEtran}
\bibliography{bibliografia}

\end{document}